\documentclass[11pt]{amsart}
\usepackage[margin=3.7cm]{geometry}
\usepackage[T1]{fontenc}
\usepackage{amsmath,amssymb,amsthm,mathtools}
\usepackage{tikz}
\usetikzlibrary{matrix,arrows,decorations.markings}
\usepackage{tikz-cd}
\usepackage{bbm}
\usepackage{array,longtable,booktabs}
\usepackage[numbers,square]{natbib}
\usepackage{url}
\usepackage{newunicodechar}
\usepackage{enumitem}
\usepackage{xcolor}
\definecolor{linkblue}{RGB}{0,82,204}
\colorlet{linkred}{red!90!blue}
\usepackage[colorlinks,linkcolor=linkred,citecolor=black,urlcolor=linkblue]{hyperref}
\newcommand{\lc}[1]{\url{#1}}
\newcommand{\lcode}{\begingroup\catcode`\_=12\relax\lcodeInner}
\newcommand{\lcodeInner}[1]{\texttt{#1}\endgroup}
\newunicodechar{¬}{\ensuremath{\lnot}}
\newunicodechar{Σ}{\ensuremath{\Sigma}}
\newunicodechar{Φ}{\ensuremath{\Phi}}
\newunicodechar{Θ}{\ensuremath{\Theta}}
\newunicodechar{Ω}{\ensuremath{\Omega}}
\newunicodechar{ᵒ}{\textsuperscript{o}}
\newunicodechar{ᵖ}{\textsuperscript{p}}
\newunicodechar{ℤ}{\ensuremath{\mathbb{Z}}}
\newunicodechar{ℕ}{\ensuremath{\mathbb{N}}}
\newunicodechar{→}{\ensuremath{\to}}
\newunicodechar{↔}{\ensuremath{\leftrightarrow}}
\newunicodechar{∀}{\ensuremath{\forall}}
\newunicodechar{∃}{\ensuremath{\exists}}
\newunicodechar{∈}{\ensuremath{\in}}
\newunicodechar{∉}{\ensuremath{\notin}}
\newunicodechar{∧}{\ensuremath{\wedge}}
\newunicodechar{∨}{\ensuremath{\vee}}
\newunicodechar{≃}{\ensuremath{\simeq}}
\newunicodechar{≅}{\ensuremath{\cong}}
\newunicodechar{≌}{\ensuremath{\backsimeq}}
\newunicodechar{≠}{\ensuremath{\neq}}
\newunicodechar{≤}{\ensuremath{\leq}}
\newunicodechar{≥}{\ensuremath{\geq}}
\newunicodechar{≫}{\ensuremath{\gg}}
\newunicodechar{≪}{\ensuremath{\ll}}
\newunicodechar{⊂}{\ensuremath{\subset}}
\newunicodechar{⊆}{\ensuremath{\subseteq}}
\newunicodechar{⊔}{\ensuremath{\sqcup}}
\newunicodechar{⊤}{\ensuremath{\top}}
\newunicodechar{⊥}{\ensuremath{\bot}}
\newunicodechar{⋙}{\ensuremath{\ggg}}
\newunicodechar{⟨}{\ensuremath{\langle}}
\newunicodechar{⟩}{\ensuremath{\rangle}}
\newunicodechar{⟦}{\ensuremath{\llbracket}}
\newunicodechar{⟧}{\ensuremath{\rrbracket}}
\newunicodechar{⟶}{\ensuremath{\longrightarrow}}
\newunicodechar{⥤}{\ensuremath{\Rightarrow}}
\newunicodechar{𝟙}{\ensuremath{\mathbbm{1}}}
\newunicodechar{ι}{\ensuremath{\iota}}
\newunicodechar{ᵢ}{\ensuremath{_i}}
\newunicodechar{₀}{\ensuremath{_0}}
\newunicodechar{₁}{\ensuremath{_1}}
\newunicodechar{₂}{\ensuremath{_2}}
\newunicodechar{₃}{\ensuremath{_3}}
\newunicodechar{₄}{\ensuremath{_4}}
\newunicodechar{α}{\ensuremath{\alpha}}
\newunicodechar{β}{\ensuremath{\beta}}
\newunicodechar{γ}{\ensuremath{\gamma}}
\newunicodechar{ν}{\ensuremath{\nu}}
\newunicodechar{ψ}{\ensuremath{\psi}}
\newunicodechar{⦃}{\{\{}
\newunicodechar{⦄}{\}\}}
\newunicodechar{⊕}{\ensuremath{\oplus}}
\newunicodechar{𝟭}{\ensuremath{\mathbbm{1}}}

\newenvironment{leancode}{%
  \begingroup
  \catcode`\_=12\relax
  \catcode`\^=12\relax
  \small\ttfamily\raggedright\baselineskip=1.15\baselineskip
  \par\smallskip\noindent
}{%
  \par\smallskip
  \endgroup
}

\theoremstyle{plain}
\newtheorem{theorem}{Theorem}[section]
\newtheorem{proposition}[theorem]{Proposition}
\newtheorem{lemma}[theorem]{Lemma}

\newtheorem{fact}[theorem]{Fact}
\theoremstyle{definition}
\newtheorem{definition}[theorem]{Definition}

\theoremstyle{remark}
\newtheorem{remark}[theorem]{Remark}

\newtheorem*{designchoice}{Design choice}

\newcommand{\calC}{\mathcal C}
\newcommand{\calD}{\mathcal D}

\newcommand{\calG}{\mathcal G}
\newcommand{\op}{\ensuremath{\mathrm{op}}}
\newcommand{\id}{\mathrm{id}}
\newcommand{\dom}{\mathrm{dom}}
\newcommand{\cod}{\mathrm{cod}}
\DeclareMathOperator{\Hom}{Hom}

\DeclareMathOperator{\Mor}{Mor}

\title[Dilatations of categories, via their Lean formalization]{Dilatations of categories,\\ via their Lean formalization}
\author{Arnaud Mayeux}
\address{University of Wisconsin--Madison, Madison, WI, USA}
\email{mayeux@wisc.edu}
\thanks{The Lean~4 source is available at \url{https://github.com/rndmx/DilCat}.}
\date{}

\begin{document}
\tolerance=2500
\emergencystretch=3em
\hbadness=4000
\maketitle

\begin{abstract}
Given a category $\calC$ and a center, that is, a family of pairs $(d_i, N_i)$ with $d_i$ a
morphism and $N_i$ a sieve over its codomain, the dilatation of $\calC$ is a new category
$\calC'$ in which every $n \in N_i$ factors, uniquely and functorially, through $d_i$; this
refines the classical localization of categories. We present the theory through a full
formalization in the Lean~4 proof assistant, on top of Mathlib: the universal property of
dilatations, the operations of restricting, shrinking, and combining centers, the dual
theory of codilatations, and the comparison with dilatations of commutative rings. The
formalization also produced minor clarifications and corrections to the published theory,
recorded in an erratum. Appendices collect a statement-by-statement dictionary to the Lean
declarations, and material on dilatations of rings relevant to the comparison with
dilatations of categories.
\end{abstract}

\setcounter{tocdepth}{1}
\tableofcontents

\section{Introduction}
\label{sec:intro}

In \citep{MayDil} we considered the benchmark problem of formalizing all indexed
mathematics as a machine-verifiable, continuously updated corpus of mathematical
knowledge, which would in particular provide large-scale data for AI-assisted
mathematics. The present paper is one step in that direction: we formalize a complete
published mathematical paper and document the formalization entirely, including the
minor clarifications and corrections that the process produced, as anticipated in
\citep[\S 5.6]{MayDil}. In this sense, the present paper is also intended as training
material for AI systems: informal statements paired with their formal counterparts,
machine-checked.

Category theory has become one of Mathlib's largest and most actively developed libraries,
and constructions once confined to informal exposition now have precise, checkable counterparts a proof
assistant can verify line by line. This paper presents the mathematics of dilatations of
categories from their Lean formalization.

\subsection{Dilatations, informally}

Let $A$ be a commutative ring and $S \subset A$ a subset. The localization $S^{-1}A$ is
obtained by formally adjoining an inverse to every element of $S$. Localization is a
pervasive operation, and its categorical analogue is equally pervasive: given a category
$\calC$ and a collection $\Sigma$ of morphisms of $\calC$, the localization $\calC[\Sigma^{-1}]$
is the category obtained by formally adjoining an inverse to every morphism in $\Sigma$,
universal among functors out of $\calC$ that invert $\Sigma$.

Dilatations of rings are a different, more refined operation. Instead of inverting an
element $a \in A$ outright, one prescribes a pair $(M,a)$ with $M$ an ideal and $a$ an
element of $A$, and forms a ring in which every element of $M$ becomes divisible by $a$,
but only those elements, and only by $a$. Dilatations of rings are
the local building blocks of dilatations of schemes (Néron blowups and their
relatives)~\citep{M,MRR,DMS}, and they specialize to localization when $M$ is taken to be the
whole ring.

The starting observation of the theory developed here is that both constructions are
instances of a single categorical construction. Fix a category $\calC$. A center in
$\calC$ is a family $\{(d_i, N_i)\}_{i \in I}$ where each $d_i$ is a morphism of $\calC$ and
each $N_i$ is a sieve over the codomain of $d_i$, i.e.\ a collection of morphisms
into $\cod(d_i)$ closed under precomposition. The dilatation of $\calC$ with center
$\{(d_i,N_i)\}_{i\in I}$ is a category $\calC'$, equipped with a functor
$\Theta : \calC \to \calC'$ that is the identity on objects, such that for every $i \in I$
and every $n \in N_i$ there is a unique morphism $b$ making the triangle
\begin{equation}
\label{eq:intro-triangle}
\begin{tikzcd}
\dom(n) \ar[rr, "\Theta(n)"] \ar[rd, dashed, "b"'] & & \cod(n) \\
& \dom(d_i) \ar[ru, "\Theta(d_i)"'] &
\end{tikzcd}
\end{equation}
commute in $\calC'$. Informally, $b$ is the ``fraction'' $d_i \backslash n = d_i^{-1}\circ n$:
$\Theta(n)$ has been made to factor through $\Theta(d_i)$. Making this triangle
condition into a genuine universal property, so that $\calC'$ is determined up to unique
isomorphism and functorial in an appropriate sense, is the technical heart of the theory,
and occupies Theorem~\ref{thm:universal-property} below.

\subsection{This paper and its formalization}

Every definition and every theorem in this text is accompanied by its Lean~4
\citep{Lean4} formalization,
carried out on top of the Mathlib library \citep{mathlib} and contained in a single file, which we refer to
throughout as the formalization. Our governing principle is that the
formalization is the primary source: where the original exposition of this material and
the Lean development differ (in how a definition is formulated, in the order in which
results are proved, in which auxiliary facts are extracted, or occasionally in the precise
hypotheses of a theorem), we follow the formalization, and we explain the discrepancy when
it is instructive. Writing a mathematical argument in
a form a proof assistant will accept forces every implicit step to be made explicit, and in
two places this exposes statements of the printed paper that are false as stated. The
corrections do not involve the main result, are technical and are easily fixable; we view
them as minor. For
\citep[Proposition~3.15]{Mayeux} (discussed in \S\ref{sec:gap} below), one step of the printed
proof is not valid; the statement is formalized here under an additional regularity
hypothesis, and a counterexample (\S\ref{sec:counterexample-315}) shows that part (2) fails
without it, and therefore also parts (3), (5) and (6), which depend on it. For \citep[Proposition~5.1]{Mayeux} (discussed in
\S\ref{sec:rings-gap} below), the printed statement is false as stated --- the formalization
exhibits an explicit counterexample --- and it is corrected by reindexing the center
(\S\ref{sec:rings-construction}), after which the intended identification holds.

\subsection{Contributions}
\label{sec:contributions}

We separate three kinds of material.
\begin{itemize}
\item \emph{Minor corrections to \citep{Mayeux}.} Theorem~\ref{thm:rings-reindex}, which
  corrects \citep[Proposition~5.1]{Mayeux}, together with the counterexample of
  \S\ref{sec:rings-gap}; and Proposition~\ref{prop:counterexample-315}, which shows that
  part (2) of \citep[Proposition~3.15]{Mayeux}, and therefore also parts (3), (5) and (6),
  which depend on it, need the additional regularity hypothesis.
\item \emph{New formalizations of known results.} All other numbered statements of
  \citep{Mayeux} listed in Appendix~\ref{app:dictionary}, with the exceptions marked there; in
  particular the universal property (Theorem~\ref{thm:universal-property},
  \citep[Theorem~3.10]{Mayeux}) and the counterexample of \citep[Fact~5.2]{Mayeux}
  (Theorem~\ref{thm:no-realizing-center}). The ring-theoretic statements of
  Appendix~\ref{app:ring-elementary} are results of \citep{M}.
\item \emph{Imported code.} The Lean construction of dilatations of rings is taken from the
  formalization \citep{MZ} (\S\ref{sec:rings-construction}); everything else in the source
  file was written for this work, on top of Mathlib.
\end{itemize}

A systematic, tabulated correspondence between every mathematical statement of this paper and
the Lean declaration(s) that formalize it is collected in Appendix~\ref{app:dictionary}, for
readers who want to locate a specific statement in the source file quickly; the body of the
paper is written to be read on its own; the dictionary is a reference, not a required
companion.

\subsection{Organization}
\label{sec:organization}

The formalization is organized, roughly, into: a reuse of Mathlib's existing machinery for
the classical localization of a category (\S\ref{sec:localization}); the definition of a
center and of the dilatation it generates, built as a quotient of a freely generated category
rather than directly as a set of equivalence classes of ``fraction sequences''
(\S\ref{sec:centers}--\S\ref{sec:dilatation-construction}); the basic properties of the
canonical functor $\Theta : \calC \to \calC'$ (\S\ref{sec:theta-properties}); the universal
property of the dilatation, which is the technical core of the theory
(\S\ref{sec:universal-property}); three constructions that produce new dilatations from old
ones, by restricting to a sub-family of the center (\S\ref{sec:restriction}), by shrinking
each sieve to a sub-sieve while keeping the same family of morphisms (\S\ref{sec:subsieve}),
or by combining two centers into one (\S\ref{sec:combining}); the dual notion of codilatation, obtained
for free by passing to the opposite category (\S\ref{sec:codilatations}); and finally two
worked examples, dilatations of commutative rings (\S\ref{sec:rings}) and an explicit
finite counterexample showing that a natural strengthening of the theory, which does hold for
rings, fails for categories (\S\ref{sec:counterexample}). We follow this order, which is the
order of the formalization, rather than the order of the original paper on this subject,
because each construction genuinely depends on the Lean infrastructure built in the sections
that precede it. Three appendices follow: the statement-by-statement dictionary
(Appendix~\ref{app:dictionary}), a short erratum to \citep{Mayeux}
(Appendix~\ref{app:erratum}), and an appendix on dilatations of rings \citep{M}, relevant to
the comparison with dilatations of categories (Appendix~\ref{app:ring-elementary}).

\subsection{Conventions}
\label{sec:conventions}

Composition is written in the mathematician's order: $g\circ f$ denotes ``$f$ then $g$''.
Mathlib, and hence every Lean snippet in this paper, composes in the opposite, diagrammatic
order: for \lcode{f : X ⟶ Y} and \lcode{g : Y ⟶ Z}, the Lean expression \lcode{f ≫ g} denotes
the composite that we write $g \circ f$. We flag this translation once here and then use it
silently throughout: a displayed equation $h = g\circ f$ always corresponds to a Lean
equation \lcode{h = f ≫ g}. For functors we also use Mathlib's diagrammatic notation
$F \ggg G$, which denotes the composite $G \circ F$.

\subsection{Availability and build instructions}
\label{sec:availability}

The formalization is the single file \lc{dilcat.lean} of the repository
\url{https://github.com/rndmx/DilCat}. The version described in this paper is the release
\texttt{v2.1} (commit \texttt{2786e7cbdf7ecdd9bdc69e0d141e4d0c6b9e8752}). It is built
with Lean \texttt{v4.34.0-rc2} and Mathlib at commit
\texttt{16edf57de7933555826307a3d81b4391a3657a09}, both pinned in the repository
(\texttt{lean-toolchain}, \texttt{lake-manifest.json}). To check it:
\begin{verbatim}
git clone https://github.com/rndmx/DilCat && cd DilCat
git checkout v2.1
lake exe cache get   # downloads the prebuilt Mathlib
lake build
\end{verbatim}
The build succeeds without errors; the file contains no \lcode{sorry} and no axiom beyond the
three standard ones (\lc{propext}, \lc{Classical.choice}, \lc{Quot.sound}), as reported by
the \texttt{\#print axioms} commands for the main theorems at the end of the file.

\section{The classical localization of a category, as reused from Mathlib}
\label{sec:localization}

\subsection{Sieves and the localization construction}

Two pieces of classical category theory are needed:
sieves, and the localization of a category at a collection of morphisms. Both are treated,
in the formalization, as already available: Mathlib provides them, and the
formalization does not reformalize either from scratch. 

A sieve $N$ over an object $Y$ of $\calC$ is a collection of morphisms with codomain
$Y$, closed under precomposition by an arbitrary morphism of $\calC$. Given a collection $E$
of morphisms of $\calC$, the sieve $S_E$ generated by $E$ consists of all composites
$e \circ n$ with $e \in E$ and $n$ an arbitrary morphism composable with $e$; when $E=\{d\}$
is a single morphism we write $S_d$ for $S_{\{d\}}$. In Lean, \lcode{Sieve X} is a structure
recording, for every object $Y$, a set of morphisms $Y \to X$ satisfying the closure
condition, and \lc{Sieve.generate} builds $S_E$ from a \lc{Presieve} (an arbitrary, not
necessarily closed, family of morphisms) representing $E$.

Given a collection $\Sigma$ of morphisms of $\calC$, the localization $\calC[\Sigma^{-1}]$ is
built classically (Gabriel--Zisman \citep{GZ}) as follows: form the directed graph $\mathcal G$ with one
vertex per object of $\calC$, one edge $a$ for every morphism $a$ of $\calC$, and one
formal inverse edge $\ell_d : \cod(d) \to \dom(d)$ for every $d \in \Sigma$; a
$\Sigma$-sequence is a finite path in $\mathcal G$; two $\Sigma$-sequences are declared
equivalent if one can be obtained from the other by composing two consecutive edges of
$\calC$, by cancelling a consecutive $d,\ell_d$ or $\ell_d,d$ pair, or by deleting an
identity edge; and $\calC[\Sigma^{-1}]$ is the category whose morphisms are equivalence
classes of $\Sigma$-sequences. This is a completely general construction, needing nothing
about $\calC$ beyond its being a category, and it satisfies the expected universal property:
a functor $F : \calC \to \calD$ that inverts every $d \in \Sigma$ factors uniquely through
the canonical functor $L : \calC \to \calC[\Sigma^{-1}]$.

This is, verbatim, the construction implemented in Mathlib's
\lc{CategoryTheory.Localization.Construction} file: the graph $\mathcal G$ is
\lc{Localization.Construction.LocQuiver}, $\Sigma$-sequences are paths in this quiver
(Mathlib's general-purpose \lc{Quiver.Path} type, via the free/path category
\lc{CategoryTheory.Paths}), the equivalence relation of \citep[Definition~2.4]{Mayeux} is
\lc{Localization.Construction.relations} (a congruence on paths in the sense of
\S\ref{sec:dilatation-construction} below), and the localization itself is presented as
\lc{(W).Localization} for \lc{W} a \lc{MorphismProperty} (Mathlib's name for a collection
of morphisms closed under no particular condition, simply a predicate
\lcode{∀ ⦃X Y⦄, (X ⟶ Y) → Prop}), together with a canonical functor \lc{W.Q} playing the
role of $L$, and the universal property is
\lc{Localization.Construction.lift}/\lc{fac}/\lc{uniq}.

\begin{designchoice}
Rather than reproduce Mathlib's localization construction adapted to $\Sigma$-sequences, the
formalization builds the raw localization $\calC[\Sigma^{-1}]$ once and defines the dilatation
as a further quotient mapping to it (\S\ref{sec:dilatation-construction}). This gives
associativity of $\Sigma$-sequence composition, and every general Mathlib fact about
localizations, for free.
\end{designchoice}

\section{Centers}
\label{sec:centers}

\subsection{The data of a center}

\begin{definition}[Center]
\label{def:center}
A center in a category $\calC$ is a family $\{(d_i,N_i)\}_{i\in I}$, indexed by a
type $I$, such that for every $i \in I$, $d_i$ is a morphism of $\calC$ and $N_i$
is a sieve over $\cod(d_i)$.
\end{definition}

As in \citep[Definition~2.9]{Mayeux}, the index type may be empty; the dilatation at the
empty center is $\calC$ itself (\lc{CatToDila_empty_isIso}). The first version of this
formalization required $I$ to be nonempty; this requirement has been removed, for centers and
for cocenters (\S\ref{sec:codilatations}), and with it the nonemptiness hypotheses on the
subsets $K \subset I$ of \S\ref{sec:restriction} and on the index type of
\lc{Center.ofMorphisms}.

A center bundles exactly the data needed to state the factorization condition of
\eqref{eq:intro-triangle}: an index set, and for each index a morphism together with a sieve
over its codomain. In Lean this is a single structure:

\begin{leancode}
structure Center (C : Type u) [Category.\{v\} C] where \\
  I : Type u \\
  dom : I → C \\
  cod : I → C \\
  mor : ∀ i : I, dom i ⟶ cod i \\
  N   : ∀ i : I, Sieve (C := C) (cod i) \\
\end{leancode}

\begin{designchoice}
Two encoding choices are worth pointing out. First, in Lean a morphism is a term of the
type \lcode{X ⟶ Y}, so its domain and codomain are part of its type, not data that can be read
off from the term. A family of morphisms with varying endpoints is therefore given by the
endpoints \lc{dom}, \lc{cod} $: I \to \calC$ together with \lcode{mor i : dom i ⟶ cod i}.
Equivalently, it could be given as a family $I \to $ \lcode{Σ X Y : C, X ⟶ Y}; the two
encodings carry the same data, and the first one lets statements mention \lcode{Z.dom i}
directly. Second, the family is
recorded as a structure with a function-valued field \lcode{I → ...} (a Mathlib-style
indexed family) rather than, say, a \lc{Set} of triples $(X,Y,f)$ with a sieve
attached to each: this is again the standard idiom for ``an $I$-indexed collection of
data'' in Lean, and it is what makes the disjoint union of two centers
(\S\ref{sec:combining}) a one-line definition, gluing two index types with
\lc{Sum}.
\end{designchoice}

We write $Z$ for a center, and, following the paper this material is drawn from, write
$\Sigma = \{d_i\}_{i\in I}$ for the underlying collection of morphisms (forgetting the
sieves). Membership in $\Sigma$ is not recorded as a \lc{Set} in Lean; instead, being a
central morphism is a property of an arbitrary triple $(X,Y,f)$:

\begin{leancode}
def IsCenterMor (f : Σ X Y : C, X ⟶ Y) : Prop := \\
  ∃ i : Z.I, f = ⟨Z.dom i, Z.cod i, Z.mor i⟩ \\[0.5\baselineskip]
def CenterMorphismProperty : MorphismProperty C := \\
  fun X Y f => IsCenterMor Z ⟨X, Y, f⟩ \\
\end{leancode}

Here \lcode{Σ X Y : C, X ⟶ Y} is Lean's dependent-pair (sigma) type: an element bundles an
object $X$, an object $Y$, and a morphism $X \to Y$ into one piece of data, giving a uniform
way to talk about ``an arbitrary morphism of $\calC$,'' regardless of its endpoints, as a
single mathematical object one can quantify over or compare for equality. \lcode{MorphismProperty C} is
Mathlib's type of predicates on morphisms of $\calC$ (formally,
\lcode{∀ ⦃X Y⦄, (X ⟶ Y) → Prop}), exactly the data needed to describe ``a collection of
morphisms'' without committing to how that collection is presented; \lc{CenterMorphismProperty}
is the \lc{MorphismProperty} cut out by $\Sigma$. Feeding this \lc{MorphismProperty} to
Mathlib's localization construction from \S\ref{sec:localization} gives, with no further
work, the raw localization $\calC[\Sigma^{-1}]$ and its canonical functor:

\begin{leancode}
def CenterLocalization : Type u := (CenterMorphismProperty Z).Localization \\[0.5\baselineskip]
def LocalizationFunctor : C ⥤ (CenterMorphismProperty Z).Localization := \\
  (CenterMorphismProperty Z).Q \\
\end{leancode}

This localization $\calC[\Sigma^{-1}]$, inverting every $d_i$ and forgetting every $N_i$, is
not the dilatation; it will however reappear constantly, in two roles: as the target of a
faithfulness hypothesis controlling uniqueness in the universal property
(\S\ref{sec:universal-property}), and, in the special case where every $N_i$ is taken to be
the largest possible sieve, as the dilatation itself:

\begin{fact}[{\citep[Fact~2.15]{Mayeux}}]
\label{fact:localization-is-dilatation}
If $N_i = S_{\id_{\cod(d_i)}}$ (the sieve of every morphism into $\cod(d_i)$) for
every $i \in I$, then the dilatation $\calC[\{(d_i)^{-1}\circ N_i\}_{i\in I}]$ is canonically
isomorphic to the raw localization $\calC[\Sigma^{-1}]$.
\end{fact}

Taking every $N_i$ maximal forces every morphism into $\cod(d_i)$ to factor through $d_i$,
which forces $d_i$ itself to become invertible: dilating by the maximal sieve is exactly
localizing. In Lean, \lc{Center.ofMorphisms} builds the center with this maximal choice of
sieve (\lcode{N := fun _ => ⊤}), and \lc{Fact_2_15} is the resulting isomorphism of
categories; \S\ref{sec:dict-5} of the appendix records how, combined with
Theorem~\ref{thm:universal-property}, this recovers \citep[Proposition~2.8]{Mayeux}'s classical universal
property of localization (\S\ref{sec:localization}) as a special case of
Theorem~\ref{thm:universal-property} rather than as a separately proved fact.

\section{Building the dilatation as a quotient of a generated category}
\label{sec:dilatation-construction}

We now fix a center $Z = \{(d_i,N_i)\}_{i\in I}$ in $\calC$ and describe how the
formalization builds the dilatation $\calC'$. 

Mathematically, $\calC'$ has a simple description: it is the subcategory of
$\calC[\Sigma^{-1}]$ with all its objects, generated by the morphisms $L(a)$ for $a$ a morphism
of $\calC$ and $L(d_i)^{-1}\circ L(n)$ for $i \in I$ and $n \in N_i$. This is the content of
\citep[Definition~2.13]{Mayeux}: a morphism of $\calC'$ is a morphism of $\calC[\Sigma^{-1}]$
admitting a representative of the shape displayed in \S\ref{sec:fraction-sequences}, so that
equality of morphisms of $\calC'$ is tested in $\calC[\Sigma^{-1}]$. A morphism of
$\calC[\Sigma^{-1}]$ admits such a representative if and only if it is a composite of the
generators above: a composite of generators is brought to that shape by absorbing each
$L(a)$ into the next numerator $n$, using that sieves are closed under precomposition. The
formal construction below presents the same category by generators and relations, and the
faithful functor of Fact~\ref{fact:faithful-into-localization} identifies it with this
subcategory: \lc{DilaGenerators} is the collection of generators above, and
\lc{DilaToLoc_range} states that a morphism of $\calC[\Sigma^{-1}]$ lies in the image of
$\calC'$ if and only if it lies in the multiplicative closure of \lc{DilaGenerators}.

\subsection{The cost of formalizing fraction sequences directly}
\label{sec:fraction-sequences}

Informally (and in the original exposition of this material), a morphism of the dilatation
$\calC'$ is an equivalence class of a $\{(d_i,N_i)\}_{i\in I}$-fraction: a
$\Sigma$-sequence in the graph $\mathcal G$ of \S\ref{sec:localization} of the special shape
\[
X_1 \xrightarrow{n_1} Y_1 \xrightarrow{\ell_{d_{i_1}}} X_2 \xrightarrow{n_2} Y_2
\xrightarrow{\ell_{d_{i_2}}} X_3 \cdots X_k \xrightarrow{n_k} Y_k
\xrightarrow{\ell_{d_{i_k}}} X_{k+1} \xrightarrow{a} X_{k+2},
\]
with $a$ an arbitrary morphism of $\calC$, $k \geq 0$, and $n_j \in N_{i_j}$ for every $j$.
Composition of such fractions, and the well-definedness of composition on equivalence
classes, are then checked directly, generalizing \citep[Fact~2.6]{Mayeux}'s computation
($\ell_{d}\circ \ell_{d'} = \ell_{d''}$ when $d,d'$ compose to a further central morphism
$d''$). Associativity is not the difficulty in this approach: once such classes are known to
be closed under composition, associativity is inherited from $\calC[\Sigma^{-1}]$
(\citep[Fact~2.11]{Mayeux}). The choice made below is one of presentation: a path category on
explicit generators comes with induction on paths of generators, and this is how existence
and uniqueness in Theorem~\ref{thm:universal-property} (\lc{generatedLocalization_commutes},
\lc{Generated_factor_unique_map}, via Lemma~\ref{lem:induction}) and the fullness statements of
\S\ref{sec:restriction} (\lc{PhiPreimage_all}, by induction on the path directly) and
\S\ref{sec:counterexample} (\lc{exists_C_mor_of_good}, via Lemma~\ref{lem:induction}) are
proved. For a compiling formalization of this approach, see \citep{MayDil} (in that paper, the formalization of dilatations of categories was an example of ongoing work, not the final formalization).

\subsection{The route taken: a freely generated category, then a quotient}

The formalization instead builds $\calC'$ in two stages, each reusing a piece of Mathlib's
generic category-theoretic infrastructure.

\textbf{Stage 1: a quiver of generators.}
\label{def:pair-fraction}
Define a quiver (a directed graph, with no
composition structure yet) on the objects of $\calC$, with two kinds of edges: one
original edge $X \to Y$ for every morphism $a : X \to Y$ of $\calC$, and one
fraction edge $\dom(n) \to \dom(d_i)$ for every $i \in I$ and every $n \in N_i$, namely the
data of a pair $(i,n)$, which we call, following \S\ref{sec:organization}'s terminology, a
$\{(d_i,N_i)\}_{i\in I}$-fraction generator. This
is exactly the data needed to write down one instance of the factorization triangle
\eqref{eq:intro-triangle}: the fraction edge for $(i,n)$ is a name for the morphism $b$ that
triangle asks for.

\begin{leancode}
def CenterSievePair : Type (max u v) := \\
  Σ i : Z.I, Σ X : C, { f : X ⟶ Z.cod i // Z.N i f } \\[0.5\baselineskip]
inductive GeneratorMorphismData (Z : Center C) {X Y : (CenterMorphismProperty Z).Localization} \\
    (f : X ⟶ Y) : Type (max u v) \\
  | fraction : PairMorWitness Z f → GeneratorMorphismData Z f \\
  | original : OriginalWitness Z f → GeneratorMorphismData Z f \\[0.5\baselineskip]
def GeneratorQuiver : Quiver (CenterMorphismProperty Z).Localization where \\
  Hom X Y := Σ f : X ⟶ Y, GeneratorMorphismData Z f \\
\end{leancode}
A \lc{CenterSievePair} bundles an index $i$, an object $X$, and a witness that some
$n : X \to \cod(d_i)$ lies in $N_i$, precisely the data $(i,n)$ indexing a fraction edge.
The quiver itself is built, slightly indirectly, on the objects of the raw
localization $\calC[\Sigma^{-1}]$ rather than directly on the objects of $\calC$ (the two
are in canonical bijection, via Mathlib's \lc{objEquiv}, since localization is the identity
on objects); this lets both kinds of generator edges (an original morphism of $\calC$, or
a fraction witnessed by a \lc{CenterSievePair}) be compared for equality as
morphisms of the raw localization, which is exactly what is needed a few lines below to
formulate the congruence of Stage~2. \lcode{GeneratorMorphismData Z f} records why a
given morphism $f$ of the raw localization arises as a generator: either it is (the image
of) an original morphism of $\calC$ (constructor \lc{original}), or it is (the image of) a
fraction-witness composite $\ell_{d_i} \circ n$ built from a \lc{CenterSievePair}
(constructor \lc{fraction}).

That underlying composite is not left informal: it is constructed in three named steps,
descending through the levels of Mathlib's localization construction
(\S\ref{sec:localization}) --- first as a path in the path category of the localization
quiver, then as its class in the raw localization.

\begin{leancode}
def inv_in_path (p : CenterSievePair Z) : \\
    ιPaths (CenterMorphismProperty Z) (Z.cod p.1) ⟶ ιPaths (CenterMorphismProperty Z) (Z.dom p.1) := \\
  Localization.Construction.ψ₂ (CenterMorphismProperty Z) (Z.mor p.1) ⟨p.1, rfl⟩ \\[0.5\baselineskip]
def fraction_in_path_single (p : CenterSievePair Z) : \\
    ιPaths (CenterMorphismProperty Z) (p.2.1) ⟶ ιPaths (CenterMorphismProperty Z) (Z.dom p.1) := \\
  Localization.Construction.ψ₁ (CenterMorphismProperty Z) p.2.2.1 ≫ inv_in_path Z p \\[0.5\baselineskip]
def fraction_in_loc_single (p : CenterSievePair Z) : \\
    objEquiv (CenterMorphismProperty Z) (p.2.1) ⟶ objEquiv (CenterMorphismProperty Z) (Z.dom p.1) := \\
  (CategoryTheory.Quotient.functor (relations (CenterMorphismProperty Z))).map (fraction_in_path_single Z p) \\
\end{leancode}

Here $\psi_1$ and $\psi_2$ are Mathlib's names (in \lc{Localization.Construction}) for the
length-one path on an original edge, respectively on a formal-inverse edge, of the
localization quiver \lc{LocQuiver} of \citep[Definition~2.2]{Mayeux}. So
\lcode{inv_in_path Z p} is (the length-one path on) the edge $\ell_{d_i}$ for the inverted
morphism $d_i$, and \lcode{fraction_in_path_single Z p} is the path-level composite
$\ell_{d_i} \circ n$ (in diagrammatic order, $\psi_1(n)$ followed by \lc{inv_in_path}).
Applying the quotient functor of the raw localization yields
\lcode{fraction_in_loc_single Z p}, the class of this composite in $\calC[\Sigma^{-1}]$;
this is exactly the morphism that \lc{IsPairMor} and \lc{PairMorWitness} --- and through
them the constructor \lc{fraction} above --- compare against, and it resurfaces in
\S\ref{sec:theta-properties} as the underlying morphism of the fraction generator edge.

\textbf{Stage 2: freely generate, then quotient.} Mathlib's \lc{CategoryTheory.Paths} functor
turns any quiver into a category: the path category, whose morphisms are finite paths
of edges, with composition given by concatenation, and associativity holding by construction
(concatenation of lists is associative). Applying it to \lcode{GeneratorQuiver Z} gives a
category \lcode{GeneratedCategory Z} whose morphisms are exactly the $\{(d_i,N_i)\}$-fractions
of the informal description above, without yet imposing any relations between different
paths representing ``the same'' fraction.

\begin{leancode}
def GeneratedCategory := CategoryTheory.Paths (GeneratorObjects Z) \\
instance : Category (GeneratedCategory Z) := Paths.categoryPaths _ \\
\end{leancode}

To obtain $\calC'$, this free category must still be quotiented by the relations that make
two representative sequences of the same fraction equal, exactly the informal content of
\citep[Definition~2.4]{Mayeux}'s elementary equivalences, transported to this setting. Rather than
re-deriving these relations combinatorially, the formalization defines the congruence
directly by reference to the raw localization, whose associated equivalence relation on
sequences is already correctly set up by Mathlib (\S\ref{sec:localization}):

\begin{leancode}
def GeneratedToLocalization : GeneratedCategory Z ⥤ (CenterMorphismProperty Z).Localization := \\
  CategoryTheory.Paths.lift (forgetGenerator Z) \\[0.5\baselineskip]
def DilaRel : HomRel (GeneratedCategory Z) := \\
  fun {_ _} f g => (GeneratedToLocalization Z).map f = (GeneratedToLocalization Z).map g \\[0.5\baselineskip]
def Dila := CategoryTheory.Quotient (DilaRel Z) \\
instance : Category (Dila Z) := CategoryTheory.Quotient.category _ \\
\end{leancode}
Here \lc{forgetGenerator} is the evident prefunctor forgetting the \lc{GeneratorMorphismData}
tag and remembering only the underlying morphism of the raw localization, and
\lc{Paths.lift} turns this prefunctor into an honest functor
\lcode{GeneratedToLocalization Z}, a functor \lcode{GeneratedCategory Z} $\to$ $\calC[\Sigma^{-1}]$ out of the
free category (again by the universal property of the path-category construction: a functor
out of a free category on a quiver is the same thing as a prefunctor on that quiver). Two
paths are declared \lc{DilaRel}-related exactly when they become equal after mapping to the
raw localization, i.e., exactly when they represent the same $\Sigma$-fraction, ignoring
the extra bookkeeping of which $N_i$'s were used. Finally, \lcode{Dila Z} is defined as the
quotient category \lcode{CategoryTheory.Quotient (DilaRel Z)}, another piece of generic Mathlib
infrastructure: given any category and any relation on its \lc{Hom}-sets, \lc{Quotient}
produces the quotient category, with composition well-defined automatically because the
construction proves once and for all that the smallest congruence containing a given relation is
compatible with composition (the \lc{Congruence} typeclass, instantiated here in three lines
by reflexivity/symmetry/transitivity of equality and compatibility of \lc{Functor.map} with
composition).

\begin{designchoice}
This two-stage route (free category on a quiver of generators, then quotient by
``agreement after forgetting to the raw localization'') obtains the category structure from
two pieces of general-purpose machinery (\lc{Paths.categoryPaths}{} for the free category,
\lc{Quotient.category} for the quotient), at the cost of introducing an auxiliary category,
\lcode{GeneratedCategory Z}, that has no name in the original exposition. This auxiliary
category is not merely a bookkeeping device: since \lcode{Dila Z} is a quotient of it and
quotient functors are full onto their generators, Theorem~\ref{thm:universal-property} is
proved by induction on it, via \lc{GeneratedCategory\_morphism\_induction}.
\end{designchoice}

\subsection{The comparison with the raw localization is faithful}

Writing $\calC'$ for \lcode{Dila Z}, the functor \lcode{GeneratedToLocalization Z} used to define
\lc{DilaRel} descends, by construction, to a functor \lcode{DilaToLoc Z}, of type
$\calC' \to \calC[\Sigma^{-1}]$ (forgetting, once again, the sieve data and remembering only
which morphisms of $\calC$ were inverted).

\begin{fact}[{\citep[Fact~2.14]{Mayeux}}]
\label{fact:faithful-into-localization}
The functor \lcode{DilaToLoc Z} $: \calC' \to \calC[\Sigma^{-1}]$ is faithful.
\end{fact}

This is the first structural fact about $\calC'$, and it already illustrates the payoff of
the two-stage construction: since two morphisms of $\calC'$ agree if and only if they are
represented by paths that are \lc{DilaRel}-related, i.e.\ paths with the same image in
$\calC[\Sigma^{-1}]$ by definition of \lc{DilaRel}, faithfulness of
\lcode{DilaToLoc Z} is close to a tautology once phrased this way: given
$f, g$ with the same image, \lc{Quot.sound} (Lean's principle that provably-related
representatives of a quotient give equal classes) directly produces $f=g$ in $\calC'$. The
formalized proof is a few lines unfolding the quotient's \lc{Hom}-type as a \lc{Quot} of paths
and invoking exactly this principle. In the description of $\calC'$ as a subcategory of
$\calC[\Sigma^{-1}]$, Fact~\ref{fact:faithful-into-localization} holds by definition.

\section{The canonical functor and its universal property}
\label{sec:universal-property}

\subsection{The canonical functor and the fraction morphisms}
\label{sec:theta-properties}

Every object of $\calC$ is, tautologically, an object of $\calC' = $\lcode{Dila Z} (both
categories share the same objects as $\calC[\Sigma^{-1}]$ and as the raw generated category);
every morphism $a$ of $\calC$ gives, via the ``original'' generator edge it defines, a
morphism of $\calC'$. This assembles into a functor.

\begin{proposition}[{\citep[Proposition~3.1(i)]{Mayeux}}]
There is a canonical functor $\Theta : \calC \to \calC'$, the identity on objects, sending a
morphism $a$ of $\calC$ to the class of the length-one path on the corresponding original
generator edge.
\end{proposition}

In Lean this is \lcode{CatToDila Z}, built directly from the quiver embedding
\lc{CToGeneratorQuiver} of \S\ref{sec:dilatation-construction}, composed with the quotient
functor \lcode{Quotient.functor (DilaRel Z)}; functoriality (respecting identities and
composition) is checked by unfolding both sides in the raw localization
$\calC[\Sigma^{-1}]$, where it is immediate.

The second half of \citep[Proposition~3.1]{Mayeux} is the existence, for every $i \in I$ and $n \in N_i$, of
the fraction morphism $b$ solving the triangle \eqref{eq:intro-triangle}, and this is, by
construction, exactly what a fraction generator edge is.

\begin{proposition}[{\citep[Proposition~3.1(ii)]{Mayeux}}]
\label{prop:fraction-exists}
For every $i \in I$, $n \in N_i$, there is a unique morphism $b : \dom(n) \to \dom(d_i)$ in $\calC'$
with $\Theta(d_i) \circ b = \Theta(n)$.
\end{proposition}

\begin{leancode}
def fraction_in_dila_single (p : CenterSievePair Z) : \\
    (CatToDila Z).obj p.2.1 ⟶ (CatToDila Z).obj (Z.dom p.1) := \\
  (GeneratedToDila Z).map (Quiver.Hom.toPath (fraction_in_generated Z p)) \\
\end{leancode}
i.e.\ $b$ is, once again, literally a name for the corresponding generator edge, viewed in
$\calC'$ instead of in the free category. Unwinding the definitions, this is the top of the
ladder begun in \S\ref{def:pair-fraction}: \lcode{fraction_in_generated Z p} is the
generator edge whose underlying raw-localization morphism is
\lcode{fraction_in_loc_single Z p}, itself the class of the path-level composite
\lcode{fraction_in_path_single Z p}, i.e.\ of \ensuremath{\psi_1}\lcode{ n }\ensuremath{\gg}\lcode{ inv_in_path Z p} --- so the
chain \lc{inv_in_path} $\to$ \lc{fraction_in_path_single} $\to$ \lc{fraction_in_loc_single}
$\to$ \lc{fraction_in_generated} $\to$ \lc{fraction_in_dila_single} consists of successive
reinterpretations of the single composite $\ell_{d_i} \circ n$, ending in $\calC'$. The defining equation
$\Theta(d_i) \circ b = \Theta(n)$ (\lc{fraction_in_dila_comp_mor}) is proved by
\lc{Quotient.sound}, reducing to a one-line computation in the raw localization exactly as in
Fact~\ref{fact:faithful-into-localization}. Uniqueness of $b$ follows from
Proposition~\ref{prop:bimorphism} below: $\Theta(d_i)$ is a monomorphism, so
$\Theta(d_i)\circ b = \Theta(n) = \Theta(d_i)\circ b'$ implies $b = b'$
(\lc{fraction_in_dila_unique}).

Recall that a bimorphism is a morphism that is both mono and epi.

\begin{fact}[{\citep[Fact~3.2]{Mayeux}}]
\label{fact:bimorphism-faithful}
The image of a morphism under a faithful functor being a bimorphism forces the morphism
itself to be a bimorphism.
\end{fact}

This is a general categorical fact, with a two-line proof by cancellation (if $F(f)$ is epi
and $a\circ f = b\circ f$ then $F(a)\circ F(f) = F(b)\circ F(f)$, so $F(a)=F(b)$ by
cancelling the epi $F(f)$, so $a=b$ by faithfulness; dually for mono), formalized as
\lc{Fact_3_2} for an arbitrary faithful $F$, not specialized to $\Theta$.

\begin{proposition}[{\citep[Proposition~3.3]{Mayeux}}]
\label{prop:bimorphism}
For every $i \in I$, $\Theta(d_i)$ is a bimorphism in $\calC'$.
\end{proposition}

Indeed \lcode{DilaToLoc Z} is faithful (Fact~\ref{fact:faithful-into-localization}) and sends
$\Theta(d_i)$ to an isomorphism of $\calC[\Sigma^{-1}]$ (localizations invert every generator
by construction), and isomorphisms are bimorphisms; Fact~\ref{fact:bimorphism-faithful} concludes.

\subsection{Sieve inclusion and $\Sigma$-regularity}
\label{sec:sigma-regular}

Two more ingredients are needed before the universal property can be stated: a sieve
inequality tracking how $\Theta$ interacts with the sieves $N_i$, and a faithfulness
condition on the target functor $F$ playing the role that faithfulness of
\lcode{DilaToLoc Z} plays for $\Theta$ itself.

For $i \in I$, write $S_{\Theta(N_i)}$ for the sieve over $\Theta(\cod(d_i))$ generated by
$\{\Theta(n) \mid n \in N_i\}$, and $S_{\Theta(d_i)}$ for the sieve generated by the single
morphism $\Theta(d_i)$.

\begin{proposition}[{\citep[Proposition~3.5]{Mayeux}}]
\label{prop:sieve-inclusion}
For every $i \in I$, $S_{\Theta(N_i)} \subset S_{\Theta(d_i)}$.
\end{proposition}

This says precisely that every morphism of $S_{\Theta(N_i)}$ (a composite $\Theta(n)\circ t$
with $n \in N_i$) factors through $\Theta(d_i)$; and indeed it does, through
$b \circ t$ with $b$ the fraction morphism of Proposition~\ref{prop:fraction-exists}. In Lean,
\lc{CatToDila_image_sieve_le_singleton} proves this by producing this explicit factorization:
unfolding membership in a pushforward sieve (Mathlib's \lc{Sieve.functorPushforward}, used
here to express $S_{\Theta(N_i)}$ as the pushforward of $N_i$ along $\Theta$) down to a
witness $(Y,h,g,hg,\text{rfl})$, and exhibiting \lc{fraction_in_dila_single}$(i,h) \circ g$ as
the required factorization, via the \lc{calc} chain
\[
\Theta(d_i) \circ (b \circ g) = (\Theta(d_i) \circ b) \circ g = \Theta(h) \circ g.
\]

\begin{definition}[{\citep[Definition~3.6]{Mayeux}}]
\label{def:sigma-regular}
A functor $F : \calC \to \calD$ is $\Sigma$-regular if the canonical functor
$\calD \to \calD[F(\Sigma)^{-1}]$ is faithful.
\end{definition}

In Lean, $\Sigma$-regularity is recorded as a \lc{Prop}, \lcode{IsSigmaRegular Z F}, defined as
faithfulness of \lcode{(ImageCenterMorphismProperty Z F).Q}, where
\lcode{ImageCenterMorphismProperty Z F} is the collection of morphisms of $\calD$ that are
images under $F$ of some $d_i$. This is deliberately not framed as membership in a
comma category $Cat_\calC^{\Sigma\text{-reg}}$ (the original exposition's route): the
formalization never needs to talk about morphisms between $\Sigma$-regular functors,
only about individual functors being $\Sigma$-regular, so recording only the property, not
the category structure around it, avoids building infrastructure that is never used.

\begin{fact}[{\citep[Fact~3.7]{Mayeux}}]
\label{fact:theta-regular}
$\Theta : \calC \to \calC'$ is $\Sigma$-regular.
\end{fact}

The functor \lcode{DilaToLoc Z} inverts every morphism of $\Theta(\Sigma)$, since it sends
$\Theta(d_i)$ to the image of $d_i$ in $\calC[\Sigma^{-1}]$. It therefore factors as the
localization $\calC' \to \calC'[\Theta(\Sigma)^{-1}]$ followed by a functor
$\calC'[\Theta(\Sigma)^{-1}] \to \calC[\Sigma^{-1}]$. As \lcode{DilaToLoc Z} is faithful
(Fact~\ref{fact:faithful-into-localization}), the localization $\calC' \to
\calC'[\Theta(\Sigma)^{-1}]$ is faithful by the following purely categorical lemma, which is
stated once and used repeatedly:

\begin{leancode}
theorem faithful_of_comp_faithful \\
    (p : C₁ ⥤ C₂) (e : C₂ ⥤ C₃) (hfaith : (p ⋙ e).Faithful) : \\
    p.Faithful := by \\
  constructor \\
  intro X Y f g h \\
  apply hfaith.map_injective \\
  simp only [Functor.comp_map, h] \\
\end{leancode}

\begin{designchoice}
This lemma (``a functor that factors, on the target side, through a faithful functor is
itself faithful'') is the single fact driving both \citep[Fact~3.7]{Mayeux} above and the following
fact, which does not otherwise appear as a named result in this section but is used
repeatedly in \S\ref{sec:combining}:
\begin{quote}
If $\Sigma' \subset \Sigma$ and the localization $\calC \to \calC[\Sigma^{-1}]$ is faithful,
then $\calC \to \calC[\Sigma'^{-1}]$ is also faithful.
\end{quote}
The original exposition proves this by a triangle-of-functors argument specific to
localizations ($\calC[\Sigma^{-1}] \cong \calC[\Sigma'^{-1}][\dots]$, then cancel). The
formalization instead isolates the one-line categorical fact both arguments really rest on,
and proves it once, in a universe-polymorphic form (\lc{faithful_of_comp_faithful_gen}) so
that it applies uniformly regardless of which universes the three categories involved happen
to live in.
\end{designchoice}

\subsection{The universal property}
\label{sec:universal-property-statement}

We can now state the theorem that makes $\calC'$ the dilatation of $\calC$ with center $Z$,
rather than merely a category equipped with a functor satisfying the factorization
triangle \eqref{eq:intro-triangle}.

\begin{theorem}[{\citep[Theorem~3.10]{Mayeux}}]
\label{thm:universal-property}
Let $F : \calC \to \calD$ be a functor such that
\begin{enumerate}
\item[(1)] for every $i \in I$, $S_{F(N_i)} \subset S_{F(d_i)}$ (the sieve generated by
  $\{F(n) \mid n\in N_i\}$ is contained in the sieve generated by $F(d_i)$ alone), and
\item[(2)] $F$ is $\Sigma$-regular (Definition~\ref{def:sigma-regular}).
\end{enumerate}
Then there is a unique functor $F' : \calC' \to \calD$ with $F' \circ \Theta = F$.
\end{theorem}

Condition (1) provides, for every fraction generator of $\calC'$, a candidate image in
$\calD$. Condition (2) is used in both halves of the proof: it makes each $F(d_i)$ a
monomorphism, so that the candidate image is unique, and it is used again to show that the
resulting functor respects the relations of $\calC'$ (\S\ref{sec:existence}). Condition (1) is
necessary for the existence of $F'$ (the forward direction of
Proposition~\ref{prop:representability}, whose proof does not use condition (2)); we do not claim
that condition (2) is necessary. In Lean:

\begin{leancode}
theorem Dila_universal_property \\
    (F : C ⥤ D) \\
    (hfaith : (ImageCenterLocalizationFunctor Z F).Faithful) \\
    (hsieve : ∀ (i : Z.I), \\
      Sieve.functorPushforward F (Z.N i) ≤ \\
        Sieve.generate (Presieve.singleton (F.map (Z.mor i)))) : \\
    ∃! (G : Dila Z ⥤ D), CatToDila Z ⋙ G = F \\
\end{leancode}
with \lc{hfaith} phrased, as in Definition~\ref{def:sigma-regular}, via the auxiliary
localization \lcode{ImageCenterLocalizationFunctor Z F} (the localization of $\calD$ at the
images of the $d_i$'s under $F$), and \lc{hsieve} the Lean transcription of condition~(1),
using \lc{Sieve.functorPushforward} for ``the sieve generated by the image of $N_i$.''

\subsubsection*{Universes}
In Lean, a category $\calC$ is given by a type of objects \lcode{C : Type u} and hom-types in a
universe \lc{v}. Mathlib's localization of $\calC$ has its objects in the same universe
\lc{u}, and so do the generated category and the dilatation \lcode{Dila Z}. The target $\calD$
of Theorem~\ref{thm:universal-property} is declared as \lcode{D : Type u'} with
\lcode{[Category.\{v'\} D]}, where \lc{u'} and \lc{v'} are arbitrary universes. In the first
version of the formalization, $\calD$ was required to have its objects in the universe \lc{u}
of the objects of $\calC$. That excluded, for instance, functors from a small category
$\calC$ to \lcode{Type u}, whose objects live in \lcode{Type (u+1)}; this case is now covered,
and is checked by an \lcode{example} in the source file. The index type of a center lives in
the universe of the objects of its category (\lcode{I : Type u}). Consequently the pushforward
of a center along a functor $F : \calC \to \calD$ (\lc{Center.pushforward},
\S\ref{sec:combining-construction}) requires the objects of $\calD$ to live in the universe
\lc{u}, which is the case in its only use, $\calD = \calC_Z$.

\begin{designchoice}
The theorem is stated with (1) and (2) as two separate, named hypotheses, rather than as
``$F$ is an object of $Cat_\calC^{\Sigma\text{-reg}}$ satisfying the sieve inequalities,''
continuing the choice already made in Definition~\ref{def:sigma-regular} not to formalize
the comma category itself. Nothing is lost: an $F$ satisfying both hypotheses is exactly an
object of $Cat_\calC^{\Sigma\text{-reg}}$ satisfying the sieve inequalities, and every use of
the theorem below supplies the two hypotheses directly rather than by first constructing a
comma-category object and then checking the sieve inequalities for it.
\end{designchoice}

The proof splits, as it must, into existence and uniqueness, and (this is the payoff of
\S\ref{sec:dilatation-construction}'s two-stage construction) both halves are organized
around the same induction principle, which we state first.

\begin{lemma}[Induction on the generated category]
\label{lem:induction}
Let $P$ be a property of morphisms of \lcode{GeneratedCategory Z}. If $P$ holds of every
identity, is closed under composition, and holds of every single generator edge, then $P$
holds of every morphism.
\end{lemma}

\begin{leancode}
lemma GeneratedCategory_morphism_induction \\
    (P : ∀ {X Y : GeneratedCategory Z}, (f : X ⟶ Y) → Prop) \\
    (h_id : ∀ X, P (𝟙 X)) \\
    (h_comp : ∀ {X Y W} (f : X ⟶ Y) (g : Y ⟶ W), P f → P g → P (f ≫ g)) \\
    (h_gen : ∀ {A B : GeneratorObjects Z} (g : (GeneratorQuiver Z).Hom A B), \\
      P (Quiver.Hom.toPath g)) : \\
    ∀ {X Y : GeneratedCategory Z} (f : X ⟶ Y), P f \\
\end{leancode}
This is Mathlib's own induction principle for the free category on a quiver
(\lc{CategoryTheory.Paths.induction}: every path is empty, or a shorter path plus one edge),
restated via $\ggg$ instead of Mathlib's internal \lc{cons} representation. Since
\lcode{GeneratorQuiver Z}'s edges are exactly the two kinds of generators of
\S\ref{sec:dilatation-construction}, an equivalent restatement of Lemma~\ref{lem:induction} is:
to check a property closed under composition and holding on identities, it suffices to
check it on every $\Theta(a)$, $a \in \Mor\calC$, and on every fraction morphism $b$. This
restatement is what is actually invoked at each of the three call sites below.

\subsubsection{Existence}
\label{sec:existence}

The construction of $F'$ proceeds generator by generator. On an original generator, there is
only one sensible choice: send $\Theta(a) \mapsto F(a)$. On a fraction generator $(i,n)$,
condition~(1) says $F(n)$ lies in the sieve $S_{F(d_i)}$, i.e.\ $F(n) = F(d_i) \circ q$ for
some $q$; condition~(2) (via \citep[Fact~3.2]{Mayeux}, applied to the faithful localization functor of
condition~(2)) makes $F(d_i)$ a bimorphism, hence $q$ is unique. This local,
generator-by-generator choice is fixed first:

\begin{leancode}
def uniqueFactor_D (hfaith : ...) (hsieve : ...) (i : Z.I) (Y : C) \\
    (n : Y ⟶ Z.cod i) (hn : Z.N i n) : F.obj Y ⟶ F.obj (Z.dom i) := \\
  Classical.choose (exists_unique_factor_D Z F hfaith hsieve i (F.obj Y) (F.map n) ...) \\
\end{leancode}
and then assembled into a prefunctor \lc{Gq} on \lcode{GeneratorQuiver Z} (sending an original
edge to $F(a)$ and a fraction edge $(i,n)$ to \lc{uniqueFactor_D} applied to $(i,n)$), which
Mathlib's \lc{Paths.lift} turns into an honest functor
$H : $\lcode{GeneratedCategory Z}$ \to \calD$ out of the free category, the same universal
property of free categories used to build \lc{GeneratedToLocalization} in
\S\ref{sec:dilatation-construction}, now used to build $H$ instead.

It remains to descend $H$ along the quotient by \lc{DilaRel}, i.e., to check that $H$
sends \lc{DilaRel}-related paths to equal morphisms of $\calD$, and this is where
condition~(2) is used a second time, more globally. The key intermediate step is:
\begin{quote}
$H$ followed by the localization $\calD \to \calD[F(\Sigma)^{-1}]$ agrees, as a
functor, with \lcode{GeneratedToLocalization Z} followed by the comparison functor
$\calC[\Sigma^{-1}] \to \calD[F(\Sigma)^{-1}]$ induced by $F$.
\end{quote}
This is \lc{generatedLocalization_commutes}, proved by Lemma~\ref{lem:induction}: on
identities and composites it is immediate; on an original generator it reduces to the
functoriality of the comparison functor; and on a fraction generator $(i,n)$ it reduces,
after unwinding both sides, to a cancellation against the (now invertible, in the further
localization) image of $F(d_i)$. Given this compatibility, two
\lc{DilaRel}-related paths $f,g$ (i.e.\ with the same image in $\calC[\Sigma^{-1}]$) have the
same image under $H$ followed by localization; since that localization is faithful
(condition~(2)), $H(f) = H(g)$ in $\calD$ outright. This descent condition is recorded as a
standalone lemma, \lc{H_descends}, and Mathlib's \lc{CategoryTheory.Quotient.lift} then
descends $H$ to a named functor --- the formalization's incarnation of $F'$ itself:

\begin{leancode}
def DilaLift (hfaith : ...) (hsieve : ...) : Dila Z ⥤ D := \\
  CategoryTheory.Quotient.lift (DilaRel Z) (H Z F hfaith hsieve) \\
    (fun _ _ f g hfg => H_descends Z F hfaith hsieve f g hfg) \\
\end{leancode}
with \lc{GeneratedToDila}$ \ggg $\lc{DilaLift}$ = H$ by construction; unwinding definitions
on generators recovers the defining equation $F' \circ \Theta = F$, recorded as
\lc{DilaLift_fac}. Because $F'$ is thus a structural definition (a quotient-lift of a
path-lift) rather than a witness extracted from an existential, it is directly reusable
downstream: the comparison functors \lc{restrictPhi}, \lc{Phi315}, and \lc{Alpha'315} of
\S\ref{sec:restriction} and \S\ref{sec:combining} are literally \lc{DilaLift} applied to the
relevant $F$, with their defining equations given by \lc{DilaLift_fac} rather than
re-extracted from an existential at each use.

\begin{designchoice}
Condition~(2) is needed already in the proof of existence, not only of uniqueness:
producing $G:\calC'\to\calD$ requires knowing $H$ respects the quotient relation ``agrees
after further localizing,'' itself a faithfulness statement about the localization on the
$\calD$ side.
\end{designchoice}

\subsubsection{Uniqueness}

Suppose $G_1, G_2 : \calC' \to \calD$ both satisfy $\Theta \ggg G_i = F$. They agree on
objects (both restrict to $F$ on objects of $\calC$, and every object of $\calC'$ is
$\Theta(X)$ for some $X$). To show they agree on morphisms, Lemma~\ref{lem:induction} is
invoked once more, transported along $\calC'$: since the quotient functor
\lc{GeneratedToDila} is full (a general fact about Mathlib's \lc{Quotient}
construction), every morphism of $\calC'$ is the image of some morphism of
\lcode{GeneratedCategory Z}, so it suffices to check $G_1, G_2$ agree (up to the evident
transport along the object-level equality) on images of the two kinds of generators:
\begin{itemize}
\item on $\Theta(a)$ for $a \in \Mor \calC$: immediate, both sides equal $F(a)$;
\item on a fraction morphism $b$ (for $(i,n)$): here the argument is exactly the one sketched
  in \S\ref{sec:theta-properties} for Proposition~\ref{prop:fraction-exists}'s uniqueness. From
  $\Theta(d_i) \circ b = \Theta(n)$, applying $G_j$ gives
  $G_j(\Theta(d_i)) \circ G_j(b) = G_j(\Theta(n)) = F(n)$; since $G_j(\Theta(d_i)) = F(d_i)$
  by hypothesis, and $F(d_i)$ is mono (condition~(2), via \citep[Fact~3.2]{Mayeux} again), the two equations
  $F(d_i)\circ G_1(b) = F(n) = F(d_i) \circ G_2(b)$ cancel to $G_1(b) = G_2(b)$.
\end{itemize}
This two-case check is recorded as \lc{Generated_factor_unique_map}, and
Lean's \lc{Functor.ext} (two functors are equal if they agree on objects and, up to
transport, on morphisms) finishes the proof as \lc{Dila_factor_unique}. Specialized to
\lc{DilaLift}, this gives \lc{DilaLift_unique} (any $G$ with $\Theta \ggg G = F$ equals
\lc{DilaLift}), and the Lean statement of Theorem~\ref{thm:universal-property} displayed
above, \lc{Dila_universal_property}, is then nothing more than the pair
(\lc{DilaLift_fac}, \lc{DilaLift_unique}) packaged as an $\exists!$.

\subsection{A representability corollary}

Restating Theorem~\ref{thm:universal-property} as a bijection statement is immediate and
convenient for the applications of \S\ref{sec:restriction}--\S\ref{sec:combining}.

\begin{proposition}[{\citep[Proposition~3.12]{Mayeux}}]
\label{prop:representability}
For $F$ a $\Sigma$-regular functor, there exists a (necessarily unique) $F'$ with
$F' \circ \Theta = F$ if and only if $S_{F(N_i)} \subset S_{F(d_i)}$ for every $i \in I$.
\end{proposition}

The forward direction is Proposition~\ref{prop:sieve-inclusion} pushed forward along $F'$
(\citep[Fact~3.11]{Mayeux}: pushforward of a generated sieve along a further functor composes correctly, a
routine but essential lemma about \lc{Sieve.functorPushforward} used here and, again, in
\S\ref{sec:combining}); the backward direction is Theorem~\ref{thm:universal-property}
itself. In Lean, \lc{CatToDila_represents} states exactly this ``iff,'' under the standing
assumption that $F$ is $\Sigma$-regular, which is more directly usable than the original
phrasing as a representable-functor statement ($\Theta$ represents a certain
$Cat_\calC^{\Sigma\text{-reg}} \to \mathrm{Set}$-valued functor): the two are logically
equivalent, and every place this fact is used below invokes it exactly in the ``iff'' form.

\section{Restricting a center}
\label{sec:restriction}

The remainder of the paper studies operations that produce new centers, and new dilatations,
from old ones. The simplest is restriction to a sub-family of indices.

Fix a center $Z=\{(d_i,N_i)\}_{i\in I}$ and a subset $K \subset I$. Restricting $Z$
to $K$ (forgetting every pair with $i \notin K$) gives a new center
$Z|_K = \{(d_i,N_i)\}_{i\in K}$ on the same category $\calC$, hence a new dilatation
$\calC[\{(d_i)^{-1}\circ N_i\}_{i\in K}]$. Since every $\{(d_i,N_i)\}_{i\in K}$-fraction is in
particular a $\{(d_i,N_i)\}_{i\in I}$-fraction, there is a canonical comparison functor.

\begin{proposition}[{\citep[Proposition~3.14]{Mayeux}}]
\label{prop:restriction}
There is a canonical functor $\Phi : \calC[\{(d_i)^{-1}\circ N_i\}_{i\in K}] \to
\calC[\{(d_i)^{-1}\circ N_i\}_{i\in I}]$. Moreover:
\begin{enumerate}
\item[(i)] \label{prop:restriction-full} if $N_i = S_{d_i}$ for every $i \in I \smallsetminus K$, then $\Phi$ is full;
\item[(ii)] if $\calC[\Gamma^{-1}] \to \calC[\Sigma^{-1}]$ is faithful, where
  $\Gamma = \{d_i\}_{i\in K}$, then $\Phi$ is faithful.
\end{enumerate}
\end{proposition}

In Lean, \lcode{Center.restrict Z K} builds $Z|_K$ (reindexing every field of \lc{Center} along
the inclusion $K \hookrightarrow I$), and $\Phi$ itself, \lc{restrictPhi}, is produced not by
hand but as \lc{DilaLift} --- the functor $F'$ of Theorem~\ref{thm:universal-property}
(\S\ref{sec:universal-property}) --- applied to
$F = \Theta_I : \calC \to \calC[\{(d_i)^{-1}\circ N_i\}_{i\in I}]$ restricted along $Z|_K$: the
sieve inequalities of condition~(1) for $Z|_K$ are exactly the $i \in K$ instances of the
sieve inequalities already known to hold for $\Theta_I$ on the whole of $Z$
(Proposition~\ref{prop:sieve-inclusion}), and condition~(2) is
Fact~\ref{fact:theta-regular} for $\Theta_I$, restricted along $K \subset I$ using the generic
faithfulness-restriction lemma of \S\ref{sec:sigma-regular}. 

\subsection{Fullness: an explicit preimage construction}

Part~(i) needs more than the universal property alone: it requires producing, for an
arbitrary morphism of $\calC[\{(d_i)^{-1}\circ N_i\}_{i\in I}]$, an explicit preimage under
$\Phi$. The proof is again organized by Lemma~\ref{lem:induction}, now applied inside the
target category's generated presentation rather than the source's: every morphism of
the big dilatation is (the image of) a path built from original and fraction generators, and
a preimage is constructed by induction on this path.

\begin{leancode}
def PhiPreimage (Z : Center C) (K : Set Z.I) \\
    {A' B' : GeneratedCategory Z} (p : A' ⟶ B') : Prop := \\
  ∃ f : ..., ...
\end{leancode}
(the precise statement records the existence of a morphism of the small dilatation mapping,
under $\Phi$, to the image of $p$). The three structural cases of the induction are handled
by \lc{PhiPreimage_id} and \lc{PhiPreimage_comp} (trivial: the identity and composition of
preimages are preimages of the identity and the composition --- the identity case even closes
by \lc{rfl}, since \lc{restrictPhi}, being the structural \lc{DilaLift} rather than an opaque
existential witness, makes the object identification $\Phi(\Theta_{Z|_K}(X)) = \Theta_I(X)$
definitional, so the \lc{eqToHom} transports flanking the identity reduce to identities), and
the generator case is where
hypothesis~(i) is used, in \lc{PhiPreimage_edge}: an original generator $\Theta_I(a)$ is
already in the image of $\Phi$ (take $\Theta_K(a)$); and a fraction generator for $(i,n)$ is
handled by cases on whether $i \in K$ (again already in the image) or $i \notin K$, in which
case hypothesis~(i) gives $n = d_i \circ q$ for some $q$, and the fraction morphism for
$(i,n)$ collapses, under $\Theta_I$, to $\Theta_I(q)$, an original generator, hence again
manifestly in the image of $\Phi$. Assembling the three cases by induction on the path
gives \lc{PhiPreimage\_all}, from which fullness
of $\Phi$, \lc{restrictPhi_full}, follows by transporting an arbitrary morphism of the
quotient back to a path (using that \lc{GeneratedToDila} is full, as in
\S\ref{sec:universal-property-statement}), applying \lc{PhiPreimage\_all}, and pushing the
resulting preimage back down.

\subsection{Faithfulness}

Part~(ii) is more direct: both dilatations map faithfully into their respective raw
localizations (Fact~\ref{fact:faithful-into-localization}), and $\Phi$ composed with the map
to the big raw localization factors through the small raw localization (essentially by
construction of $\Phi$ via the universal property); hypothesis~(ii)'s faithfulness of the
comparison between the two raw localizations then gives faithfulness of $\Phi$ itself via
\lc{faithful_of_comp_faithful}, the same one-line categorical lemma from
\S\ref{sec:sigma-regular} used for \citep[Fact~3.7]{Mayeux} and, again, below.

\section{Shrinking the sieves}
\label{sec:subsieve}

A second comparison functor sits alongside Proposition~\ref{prop:restriction}'s $\Phi$:
rather than dropping indices from $I$, one can shrink each sieve $N_i$ to an arbitrary
sub-sieve $M_i \subset N_i$, keeping the same family of morphisms $\{d_i\}_{i\in I}$.

\begin{fact}[{\citep[Fact~3.13]{Mayeux}}]
\label{fact:subsieve}
Let $Z=\{(d_i,N_i)\}_{i\in I}$ be a center and $M_i \subset N_i$ a sieve over $\cod(d_i)$ for
every $i$; write $Z_M$ for the center $\{(d_i,M_i)\}_{i\in I}$. There is a canonical functor
$\varphi : \calC[\{(d_i)^{-1}\circ M_i\}_{i\in I}] \to \calC'$, and $\varphi$ is faithful.
\end{fact}

As with $\Phi$ in \S\ref{sec:restriction}, $\varphi$ (\lc{Fact313Phi}) is built
directly from Theorem~\ref{thm:universal-property} applied to $\Theta : \calC \to \calC'$:
$\Sigma$-regularity of $\Theta$ transports unchanged to $Z_M$ (\lc{IsSigmaRegular_altSieve},
since $Z_M$ shares the same underlying $\Sigma$ as $Z$), and the sieve inequality is the
subsieve inclusion $M_i \subset N_i$ composed with Proposition~\ref{prop:sieve-inclusion}'s
inequality for $Z$. Faithfulness is where the shared-$\Sigma$ observation pays off a second
time: since $Z_M$ and $Z$ have the same $\Sigma$, they have the same raw localization
and the same functor $L=$\lc{LocalizationFunctor}, so $\varphi \ggg$\lc{DilaToLoc}$\ Z$ and
\lc{DilaToLoc}$\ Z_M$ both factor $L$ through $\Theta_{Z_M}$, so \lc{Dila_factor_unique}
identifies them outright, and \lc{DilaToLoc}$\ Z_M$ is faithful unconditionally
(Fact~\ref{fact:faithful-into-localization}), so \lc{faithful_of_comp_faithful} gives
faithfulness of $\varphi$. 

\begin{designchoice}
The printed proof is one sentence (an $M$-fraction is already an $N$-fraction, so $\varphi$ is
the identity on representatives); that does not transport here, since the two dilatations are
quotients of different generated categories, so $\varphi$ is instead built from
Theorem~\ref{thm:universal-property} and \lc{Dila_factor_unique}.
\end{designchoice}

\section{Two isomorphisms between dilatations built in stages from different centers}
\label{sec:combining}

This section is about two procedures, closely related but genuinely distinct. Both combine
two pieces of data on the same category $\calC$ into one and compare the resulting dilatation
with a dilatation carried out in two stages: first dilate by one, then dilate the result by
(the pushforward of) the other.

In the first (\S\ref{sec:combining-construction}--\S\ref{sec:gap}) the two pieces of data are
two centers, with different generators; this is \citep[Proposition~3.15]{Mayeux}. We first
recall the printed statement, then describe the construction and the declarations it
produces, in the terms the source file uses, and then compare the two. The printed statement
and the formal one do not say the same thing: the formal one carries an additional regularity
hypothesis, which cannot be removed (\S\ref{sec:counterexample-315}).

In the second (\S\ref{sec:combining-sieves}) the two pieces of data share their generators and
differ only in their sieves; this is \citep[Proposition~3.18]{Mayeux}, and it is formalized
unconditionally.

\subsection{Partial dilatation, then partial dilatation, equals the full dilatation}
\label{sec:combining-construction}

Fix two centers $Z = \{(d_i,N_i)\}_{i\in I}$ and $W = \{(d_j,N_j)\}_{j\in J}$ on the same
category $\calC$. Write $\Theta_Z : \calC \to \calC_Z$ and
$\Theta_{I'} : \calC \to \calC_{I'}$ for the canonical functors, $\beta : \calC_Z \to
\calC_Z[\{\Theta_Z(d_j)^{-1}\circ\Theta_Z(N_j)\}_{j\in J}]$ for the canonical functor of the
dilatation of $\calC_Z$ by the pushed-forward center $W$, and $\Phi : \calC_Z \to
\calC_{I'}$ for the comparison functor of \citep[Proposition~3.14]{Mayeux}. The statement as
printed in \citep[Proposition~3.15]{Mayeux} is:
\begin{quote}
With $\calC_Z := \calC[\{(d_i)^{-1}\circ N_i\}_{i\in I}]$,
$\calC_{I'} := \calC[\{(d_k)^{-1}\circ N_k\}_{k\in I'}]$ for $I' = I\sqcup J$, and
$\calC_Z[\{\Theta_Z(d_j)^{-1}\circ\Theta_Z(N_j)\}_{j\in J}]$ the dilatation of $\calC_Z$ by the
pushed-forward $W$:
\begin{enumerate}
\item[(1)] $\Phi$ is $\{\Theta_Z(d_j)\}_{j\in J}$-regular;
\item[(2)] $\Theta_Z \ggg \beta$ is $\{d_k\}_{k\in I'}$-regular;
\item[(3)] there is a unique $\alpha : \calC_{I'} \to \calC_Z[\{\Theta_Z(d_j)^{-1}\circ
  \Theta_Z(N_j)\}_{j\in J}]$ with $\Theta_Z \ggg \beta = \Theta_{I'} \ggg \alpha$;
\item[(4)] there is a unique
  $\alpha' : \calC_Z[\{\Theta_Z(d_j)^{-1}\circ \Theta_Z(N_j)\}_{j\in J}] \to \calC_{I'}$ with
  $\Phi = \beta \ggg \alpha'$;
\item[(5)] $\alpha \ggg \alpha' = \id$ and $\alpha' \ggg \alpha = \id$;
\item[(6)] hence $\calC_Z[\{\Theta_Z(d_j)^{-1}\circ \Theta_Z(N_j)\}_{j\in J}] \cong \calC_{I'}$.
\end{enumerate}
\end{quote}
Informally, this says: dilating twice, first by $Z$ and then by (the pushforward of) $W$,
gives the same category as dilating once by the union of $Z$ and $W$.

\subsubsection*{The combined center}
The formalization first forms the center that simply puts $Z$ and $W$ side by side, indexed
by the disjoint union $I \sqcup J$:
\begin{leancode}
def Center.sum (Z W : Center C) : Center C where \\
  I := Z.I ⊕ W.I \\
  dom := Sum.elim Z.dom W.dom \\
  cod := Sum.elim Z.cod W.cod \\
  mor := fun i => match i with | Sum.inl i => Z.mor i | Sum.inr j => W.mor j \\
  N   := fun i => match i with | Sum.inl i => Z.N i   | Sum.inr j => W.N j \\
\end{leancode}
every field is simply routed to the $Z$- or the $W$-side by cases on the summand. Two
immediate containment facts about the associated collections of central morphisms,
\lc{CenterMorphismProperty_sum_inl_le} and \lc{CenterMorphismProperty_sum_inr_le}
(each a two-line proof: an index witnessing $f \in \Sigma_Z$ or $f\in\Sigma_W$ certainly
witnesses $f \in \Sigma_{Z\sqcup W}$, via \lc{Sum.inl} or \lc{Sum.inr}), together with
their $F$-image analogues \lc{ImageCenterMorphismProperty_sum_inl_le} and
\lc{...\_sum_inr_le} for an arbitrary functor $F$ out of $\calC$, are recorded once and
reused repeatedly below: they are the only place the definition of \lc{Center.sum} is
unfolded directly, everything else works with these two containments as a black box.

\subsubsection*{Regularity of $\Theta_{Z\sqcup W}$ on each side}
Before anything is built, the formalization records that $\Theta_{Z\sqcup W} : \calC \to
\calC_{Z\sqcup W}$ (which is already known to be $\Sigma_{Z\sqcup W}$-regular, as an
instance of Fact~\ref{fact:theta-regular} applied to the combined center) remains regular
when restricted to either half of the index set:
\begin{leancode}
lemma CatToDila_isSigmaRegular_sum_inl (Z W : Center C) : \\
    IsSigmaRegular Z (CatToDila (Z.sum W)) := by \\
  show (ImageCenterMorphismProperty Z (CatToDila (Z.sum W))).Q.Faithful \\
  apply faithful_of_comp_faithful \\
    (ImageCenterMorphismProperty Z (CatToDila (Z.sum W))).Q \\
    (Localization.Construction.lift ... (ImageCenterMorphismProperty (Z.sum W) \\
      (CatToDila (Z.sum W))).Q ...) \\
  rw [Localization.Construction.fac] \\
  exact CatToDila_isSigmaRegular (Z.sum W) \\
\end{leancode}
with a symmetric statement \lc{CatToDila_isSigmaRegular_sum_inr} for $W$. Both are instances
of the same one-line categorical lemma from \S\ref{sec:sigma-regular},
\lc{faithful_of_comp_faithful}: the localization at the $Z$-part alone factors, via
\lc{Localization.Construction.lift}, through the localization at the full $Z\sqcup W$
family (using the containment \lc{ImageCenterMorphismProperty_sum_inl_le} above to know
every morphism the bigger localization needs to invert to make this factoring functor
well-defined is already inverted), and the bigger localization is faithful by
Fact~\ref{fact:theta-regular}. This is the same ``sub-family of a regular family is regular''
pattern already used in \S\ref{sec:restriction}, now instantiated for
the two halves of a disjoint union instead of for an arbitrary subset $K \subset I$.

\subsubsection*{$\beta$: dilating $\calC_Z$ by the pushed-forward $W$}
Pushing $W$ forward along $\Theta_Z : \calC \to \calC_Z := $\lc{Dila}$\ Z$ gives a center on
$\calC_Z$ itself, and dilating $\calC_Z$ by it gives a functor $\beta$:
\begin{leancode}
def CenterZW (Z W : Center C) : Center (Dila Z) := W.pushforward (CatToDila Z) \\
def Beta315 (Z W : Center C) : Dila Z ⥤ Dila (CenterZW Z W) := CatToDila (CenterZW Z W) \\
\end{leancode}
\lc{Center.pushforward} is the general-purpose operation transporting a center along an
arbitrary functor (sieve $N_j \mapsto$ its pushforward \lc{Sieve.functorPushforward}$\ 
(\Theta_Z)\ N_j$, already used for Proposition~\ref{prop:sieve-inclusion}); $\beta$ is then,
verbatim, $\Theta$ for this pushed-forward center, i.e.\ nothing more than another instance of
Definition~\ref{def:center}--\S\ref{sec:dilatation-construction}'s basic construction, applied
one level up, inside $\calC_Z$ instead of inside $\calC$.

\subsubsection*{$\Phi$: comparing $\calC_Z$ directly with $\calC_{Z\sqcup W}$}
A comparison functor $\Phi : \calC_Z \to \calC_{Z\sqcup W}$ is produced next, not by
invoking Proposition~\ref{prop:restriction}'s \lc{restrictPhi} (which would be the obvious
route, $Z\sqcup W$ restricted to its $Z$-part being $Z$ again) but by a fresh application of
Theorem~\ref{thm:universal-property} to $\Theta_{Z\sqcup W}$ itself:
\begin{leancode}
noncomputable def Phi315 (Z W : Center C) : Dila Z ⥤ Dila (Z.sum W) := \\
  DilaLift Z (CatToDila (Z.sum W)) \\
    (CatToDila_isSigmaRegular_sum_inl Z W) (CatToDila_sum_hsieve_inl Z W) \\[0.5\baselineskip]
lemma Phi315_spec (Z W : Center C) : \\
    CatToDila Z ⋙ Phi315 Z W = CatToDila (Z.sum W) := \\
  DilaLift_fac Z (CatToDila (Z.sum W)) \\
    (CatToDila_isSigmaRegular_sum_inl Z W) (CatToDila_sum_hsieve_inl Z W) \\
\end{leancode}

\subsubsection*{Regularity of $\Phi$}
\begin{leancode}
theorem Phi315_isSigmaRegular (Z W : Center C) : \\
    IsSigmaRegular (CenterZW Z W) (Phi315 Z W) := by \\
  show (ImageCenterMorphismProperty (CenterZW Z W) (Phi315 Z W)).Q.Faithful \\
  rw [ImageCenterMorphismProperty_ZW_Phi_eq] \\
  exact CatToDila_isSigmaRegular_sum_inr Z W \\
\end{leancode}
The single nontrivial ingredient is \lc{ImageCenterMorphismProperty_ZW_Phi_eq}, an
extensionality lemma identifying the images of the \lc{CenterZW}-generators under $\Phi$
with the images of the $W$-generators under $\Theta_{Z\sqcup W}$ directly (using
\lc{Phi315_spec} to transport along the object-level identification $\Phi(\Theta_Z(X)) =
\Theta_{Z\sqcup W}(X)$, with the \lc{eqToHom} bookkeeping this identification carries,
exactly as in \S\ref{sec:universal-property-statement}); once the two image-classes of
morphisms are known to coincide, regularity is exactly
\lc{CatToDila_isSigmaRegular_sum_inr} from above.

\subsubsection*{The comparison functor in the other direction, built first}
The formalization builds a second comparison functor,
$\alpha' : \calC_Z[$\lc{CenterZW}$] \to \calC_{Z\sqcup W}$, before it builds the one going the
other way, and the source file's own comment explains why: the construction still to come
will need this one's defining equation as an ingredient, so it is built first, and it needs
nothing beyond the regularity of $\Phi$ just established:
\begin{leancode}
noncomputable def Alpha'315 (Z W : Center C) : Dila (CenterZW Z W) ⥤ Dila (Z.sum W) := \\
  DilaLift (CenterZW Z W) (Phi315 Z W) \\
    (Phi315_isSigmaRegular Z W) (Phi315_hsieve Z W) \\[0.5\baselineskip]
theorem Alpha'315_spec (Z W : Center C) : \\
    Beta315 Z W ⋙ Alpha'315 Z W = Phi315 Z W := \\
  DilaLift_fac (CenterZW Z W) (Phi315 Z W) \\
    (Phi315_isSigmaRegular Z W) (Phi315_hsieve Z W) \\
\end{leancode}
This is Theorem~\ref{thm:universal-property} applied a third time in this section alone (after
the two applications already used for $\Phi$): to $\Phi$ itself, viewed as a functor out of
$\calC_Z$'s \lc{CenterZW}-dilatation, using exactly the regularity and sieve-inclusion facts
already in hand (\lc{Phi315_hsieve}, the pushforward-sieve inequality, is again
Proposition~\ref{prop:sieve-inclusion} transported through \lc{Sieve.functorPushforward_comp},
the same pattern as \citep[Fact~3.11]{Mayeux}).

\subsubsection*{The delicate direction, and the hypothesis it needs}
The comparison functor going the other way, $\alpha : \calC_{Z\sqcup W} \to
\calC_Z[$\lc{CenterZW}$]$, is delicate. Its defining
application of Theorem~\ref{thm:universal-property} needs $\Theta_Z\ggg\beta$ to be
$\Sigma_{Z\sqcup W}$-regular, which does not hold in general
(\S\ref{sec:counterexample-315}). The formalization takes it as an explicit hypothesis,
threaded through every following declaration by name:
\begin{leancode}
noncomputable def Alpha315 (Z W : Center C) \\
    (hreg : IsSigmaRegular (Z.sum W) (CatToDila Z ⋙ Beta315 Z W)) : \\
    Dila (Z.sum W) ⥤ Dila (CenterZW Z W) := \\
  (Dila_universal_property (Z.sum W) (CatToDila Z ⋙ Beta315 Z W) \\
      hreg (BetaComp315_hsieve Z W)).choose \\
\end{leancode}
where \lc{BetaComp315_hsieve} is the accompanying sieve-inclusion hypothesis.

\subsubsection*{Assembling the isomorphism}
With \lc{hreg} in hand, the remaining declarations are again routine, and again go through
Theorem~\ref{thm:universal-property}'s uniqueness clause, now via a standalone helper,
\lc{Dila_factor_unique}, recording exactly that clause for reuse (\S\ref{sec:universal-property-statement}), applied three times in succession. First, an auxiliary identity:
\begin{leancode}
theorem Phi315_comp_Alpha315 (Z W : Center C) \\
    (hreg : IsSigmaRegular (Z.sum W) (CatToDila Z ⋙ Beta315 Z W)) : \\
    Phi315 Z W ⋙ Alpha315 Z W hreg = Beta315 Z W := by \\
  apply Dila_factor_unique Z (CatToDila Z ⋙ Beta315 Z W) \\
    (Phi315 Z W ⋙ Alpha315 Z W hreg) (Beta315 Z W) \\
  · show CatToDila Z ⋙ Phi315 Z W ⋙ Alpha315 Z W hreg = CatToDila Z ⋙ Beta315 Z W \\
    rw [← Functor.assoc, Phi315_spec, Alpha315_spec] \\
  · rfl \\
  · exact IsSigmaRegular_sum_inl_of Z W (CatToDila Z ⋙ Beta315 Z W) hreg \\
\end{leancode}
(both $\Phi\ggg\alpha$ and $\beta$ agree after precomposing with $\Theta_Z$, so
\lc{Dila_factor_unique} (fed the restriction of \lc{hreg} to the $Z$-part alone, via
\lc{IsSigmaRegular_sum_inl_of}, the same restriction-of-regularity fact used throughout
\S\ref{sec:combining-construction}) identifies them). From this single identity, two
further identities follow by the same pattern, cancelling in the other two dilatations
in turn:
\begin{leancode}
theorem Alpha315_comp_Alpha'315 (Z W : Center C) \\
    (hreg : ...) : Alpha315 Z W hreg ⋙ Alpha'315 Z W = 𝟭 (Dila (Z.sum W)) := by \\
  apply Dila_factor_unique (Z.sum W) (CatToDila (Z.sum W)) ... (𝟭 _) \\
  · show CatToDila (Z.sum W) ⋙ Alpha315 Z W hreg ⋙ Alpha'315 Z W = CatToDila (Z.sum W) \\
    rw [← Functor.assoc, Alpha315_spec, Functor.assoc, Alpha'315_spec, Phi315_spec] \\
  · exact Functor.comp_id _ \\
  · exact CatToDila_isSigmaRegular (Z.sum W) \\[0.5\baselineskip]
theorem Alpha'315_comp_Alpha315 (Z W : Center C) \\
    (hreg : ...) : Alpha'315 Z W ⋙ Alpha315 Z W hreg = 𝟭 (Dila (CenterZW Z W)) := by \\
  apply Dila_factor_unique (CenterZW Z W) (Beta315 Z W) ... (𝟭 _) \\
  · show Beta315 Z W ⋙ Alpha'315 Z W ⋙ Alpha315 Z W hreg = Beta315 Z W \\
    rw [← Functor.assoc, Alpha'315_spec, Phi315_comp_Alpha315] \\
  · exact Functor.comp_id _ \\
  · exact CatToDila_isSigmaRegular (CenterZW Z W) \\
\end{leancode}
and finally, a pair of mutually inverse functors assembles into an isomorphism of \lc{Cat}:
\begin{leancode}
noncomputable def Iso315 (Z W : Center C) \\
    (hreg : IsSigmaRegular (Z.sum W) (CatToDila Z ⋙ Beta315 Z W)) : \\
    Cat.of (Dila (CenterZW Z W)) ≅ Cat.of (Dila (Z.sum W)) where \\
  hom := (Alpha'315 Z W).toCatHom \\
  inv := (Alpha315 Z W hreg).toCatHom \\
  hom_inv_id := Cat.ext <| Alpha'315_comp_Alpha315 Z W hreg \\
  inv_hom_id := Cat.ext <| Alpha315_comp_Alpha'315 Z W hreg \\
\end{leancode}
needing only the two composite identities just proved, not the fuller coherence data of a
\lc{CategoryTheory.Equivalence}.

\subsection{Comparison with \citep[Proposition~3.15]{Mayeux}}
\label{sec:gap}

Having built \lc{Center.sum}, $\Phi$, $\beta$, \lc{Alpha'315}, \lc{Alpha315}, and
\lc{Iso315} in the order above, we compare them with the six parts of
\citep[Proposition~3.15]{Mayeux} recalled in \S\ref{sec:combining-construction}.

Parts (1) and (4) match the construction of \S\ref{sec:combining-construction} exactly and
unconditionally: (1) is \lc{Phi315_isSigmaRegular}; (4) is \lc{Alpha'315} together with
\lc{Alpha'315_spec} and \lc{Alpha'315_unique}. Parts (3), (5), (6) match the construction too,
but only conditionally, since $\alpha$ itself already needs part~(2)'s regularity to be
built at all: (3) is \lc{Alpha315} together with \lc{Alpha315_spec} and \lc{Alpha315_unique},
each taking \lc{hreg} as an explicit argument; (5) is \lc{Alpha315_comp_Alpha'315} and
\lc{Alpha'315_comp_Alpha315}, both stated for an arbitrary \lc{hreg}; (6) is \lc{Iso315},
likewise. In other words, item~(2) is not merely ``the part where the two accounts part
ways'' in isolation; it is a hypothesis that then propagates through every part built on
top of $\alpha$, i.e.\ through (3), (5), and (6) as well.

\subsubsection*{Where the printed proof of part (2) fails}

The printed proof of part~(2) applies \citep[Fact~2.14]{Mayeux} twice, to obtain a faithful
functor from $\calC_Z[\{\Theta_Z(d_j)^{-1}\circ\Theta_Z(N_j)\}_{j\in J}]$ to
$\calC[(\{d_i\}_{i\in I})^{-1}][(\{d_j\}_{j\in J})^{-1}]$. The second application requires that
the localization at $\{\Theta_Z(d_j)\}_{j\in J}$ of the faithful functor
$\calC_Z \to \calC[(\{d_i\}_{i\in I})^{-1}]$ is again faithful, and a localization of a
faithful functor need not be faithful. In the example of \S\ref{sec:counterexample-315},
$\calC_Z = \calC \to \calC[d^{-1}]$ is faithful, while $\calC[e^{-1}] \to \calC[\{d,e\}^{-1}]$ is
not.

\subsection{A counterexample to \citep[Proposition~3.15]{Mayeux}~(2), (3), (5), (6)}
\label{sec:counterexample-315}

\begin{proposition}
\label{prop:counterexample-315}
Let $\calC$ be the category associated with the ordered set $0 < 1 < 2$, with morphisms
$u : 0 \to 1$, $d : 1 \to 2$ and $e = d\circ u : 0 \to 2$. Let $Z = \{(d, S_d)\}$ and
$W = \{(e, \top)\}$, where $S_d = \{d, e\}$ is the sieve generated by $d$ and $\top$ is the
maximal sieve over~$2$. Write $E := \calC_Z[\{\Theta_Z(e)^{-1}\circ\Theta_Z(\top)\}]$ for the
two-stage dilatation and $\calC_{I'}$ for the dilatation at $Z \sqcup W$. Then:
\begin{enumerate}
\item[(i)] $\Theta_Z \ggg \beta$ is not $\{d, e\}$-regular;
\item[(ii)] the functor $\alpha' : E \to \calC_{I'}$ of part~(4) is not faithful;
\item[(iii)] there is no functor $\alpha : \calC_{I'} \to E$ with
  $\Theta_Z \ggg \beta = \Theta_{I'} \ggg \alpha$;
\item[(iv)] $E$ and $\calC_{I'}$ are not equivalent categories.
\end{enumerate}
In particular, parts (2), (3), (5) and (6) of \citep[Proposition~3.15]{Mayeux} fail for
$\calC$, $Z$, $W$.
\end{proposition}

\begin{proof}
Every functor out of $\calC$ is faithful, since $\calC$ is thin. The fraction generators of
$Z$ are $d\backslash d = \id_1$ and $d\backslash e = u$, so the functor
$\Theta_Z : \calC \to \calC_Z$ is full, and $\calC_Z$ is identified with $\calC$. Under this
identification, $E$ is the dilatation of $\calC$ at $\{(e,\top)\}$, that is the localization
$\calC[e^{-1}]$ (Fact~\ref{fact:localization-is-dilatation}). In $E$ put
$s := u \circ e^{-1} : 2 \to 1$. Then $d \circ s = e \circ e^{-1} = \id_2$.

We show $s \circ d \neq \id_1$ in $E$. Let $G : \calC \to \mathrm{Set}$ send $0$ and $2$ to a
one-point set, $1$ to $\{a,b\}$, $u$ to the map with image $a$, and $d$ to the unique map to
the point. Then $G(e)$ is a bijection, so $G$ extends to $E$, and $G(s\circ d)$ is the
constant map with value $a$, which is not $\id_{\{a,b\}}$. Hence $s\circ d\neq\id_1$, and since
$d \circ (s \circ d) = d = d \circ \id_1$, the morphism $d$ is not a monomorphism of $E$.

(i) In any localization of $E$ inverting $d$, the equality $d \circ s = \id_2$ gives
$s \circ d = \id_1$. So the localization $E \to E[\{d,e\}^{-1}]$ identifies the distinct
morphisms $s\circ d$ and $\id_1$ and is not faithful.

In $\calC_{I'}$, the morphisms $\Theta_{I'}(d)$ and $\Theta_{I'}(e)$ are monomorphisms
(Proposition~\ref{prop:bimorphism}). The fraction morphism $c := e\backslash\id_2$ satisfies
$\Theta_{I'}(e)\circ c = \id_2$, hence
$\Theta_{I'}(e)\circ c\circ\Theta_{I'}(e) = \Theta_{I'}(e)$ and $c\circ \Theta_{I'}(e) = \id_0$:
$\Theta_{I'}(e)$ is invertible. Then $\Theta_{I'}(d) \circ (\Theta_{I'}(u)\circ c) = \id_2$,
and cancelling the monomorphism $\Theta_{I'}(d)$ in
$\Theta_{I'}(d)\circ(\Theta_{I'}(u)\circ c)\circ\Theta_{I'}(d) = \Theta_{I'}(d)$ shows that
$\Theta_{I'}(d)$ is invertible.

(ii) The functor $\alpha'$ satisfies $\alpha'(d) = \Theta_{I'}(d)$, which is invertible, so
$\alpha'(d)\circ\alpha'(s) = \id_2$ gives $\alpha'(s\circ d) = \id_1 = \alpha'(\id_1)$, while
$s \circ d \neq \id_1$.

(iii) Such an $\alpha$ would send the invertible morphism $\Theta_{I'}(d)$ to $d$, so $d$
would be invertible in $E$, hence a monomorphism.

(iv) In the localization $\calC[\{d,e\}^{-1}]$, every morphism is invertible: the images of
$d$ and $e$ are invertible by construction, the image of $u$ is invertible since
$d\circ u = e$, and the formal inverses are invertible. Since the functor
$\calC_{I'} \to \calC[\{d,e\}^{-1}]$ is faithful (Fact~\ref{fact:faithful-into-localization}),
every morphism of $\calC_{I'}$ is a monomorphism. An equivalence $E \simeq \calC_{I'}$ would
reflect monomorphisms, so $d$ would be a monomorphism of $E$, which it is not. Hence $E$ and
$\calC_{I'}$ are not equivalent (\lc{all_iso_loc}, \lc{all_mono}, \lc{not_equivalent}).

Part (2) of \citep[Proposition~3.15]{Mayeux} is (i). Part (3) fails by (iii), and with it
parts (5) and (6), since by (ii) and (iv) the comparison $\alpha'$ is not an isomorphism and
no isomorphism exists.
\end{proof}

In Lean, the example is \lc{Ex315.Obj} (the three-element order, as a thin category), with
centers \lc{Ex315.Z} and \lc{Ex315.W}; the four parts are \lc{Ex315.not_hreg},
\lc{Ex315.alpha'_not_faithful}, \lc{Ex315.no_alpha} and \lc{Ex315.not_equivalent}. The
separating functor $G$ is \lc{Ex315.G}, valued in a small category \lc{Ex315.Tgt} with two
objects, a one-point set and a two-point set, and all maps between them.

\begin{remark}[the ring-theoretic counterpart]
\label{rem:rings-counterpart}
For an element $a$ of a ring, being regular (a non-zero-divisor) is the same as injectivity
of $A \to A[a^{-1}]$, which is what Definition~\ref{def:sigma-regular} asks of the one-object
category attached to $A$; and in a dilatation the images of the $a_i$ are always
non-zero-divisors (\citep[Fact~2.10]{M}). Over rings, therefore, the regularity needed to
iterate the construction is a theorem, and the counterpart of
\citep[Proposition~3.15]{Mayeux} holds unconditionally: dilating along $K \subset I$ and then
along $I \setminus K$ recovers the one-stage dilatation
(Proposition~\ref{prop:app-2-24} in Appendix~\ref{app:ring-elementary}, formalized without any
analogue of \lc{hreg}). This is \citep[Proposition~2.24]{M}.
\end{remark}

\subsection{Comparing dilatation with the combination of sieves}
\label{sec:combining-sieves}

A close relative of the construction above combines, on the same family of morphisms
$\{d_i\}_{i\in I}$, two different choices of sieves rather than two different centers, and here
the formalization proves the analogous regularity unconditionally.

\subsubsection*{The Lean construction}
For each $i \in I$, let $N_i'$ be another sieve over $\cod(d_i)$, and $N_i'' := N_i\cup N_i'$
their union (\lc{Center.sieveUnion}); let \lc{Center.altSieve} denote the center on $\calC$
with the same generators $\{d_i\}_{i\in I}$ and sieves $N_i'$ in place of $N_i$, so that the
two-copy center carrying both sieve families side by side is the sum
\lcode{Z.sum (Z.altSieve N')} of \S\ref{sec:combining-construction}. The two comparison
functors are built exactly as \lc{Alpha315} and \lc{Alpha'315} were, via
Theorem~\ref{thm:universal-property}, applied once in each direction, and recorded as
\lc{Alpha318} and \lc{Alpha'318}, with \lc{Iso318} assembling them into an isomorphism of
categories, mirroring \lc{Iso315} verbatim.

\subsubsection*{Comparison with the printed paper}
\begin{proposition}[{\citep[Proposition~3.18]{Mayeux}}]
\label{prop:sieve-union}
With notation as above, $\calC[\{(d_i)^{-1}\circ N_i\}_{i\in I}, \{(d_i)^{-1}\circ
N_i'\}_{i\in I}]$ (the dilatation at the two-copy center carrying both families of sieves
on the same generators) is canonically isomorphic to
$\calC[\{(d_i)^{-1}\circ N_i''\}_{i\in I}]$.
\end{proposition}

\section{Codilatations}
\label{sec:codilatations}

Everything so far dilates along morphisms $d_i : \dom(d_i) \to \cod(d_i)$ using
sieves $N_i$ over $\cod(d_i)$ (collections closed under precomposition), forcing
morphisms into $\cod(d_i)$ to factor through $d_i$. There is an evidently dual
construction, forcing morphisms out of $\dom(d_i)$ to factor through $d_i$, using
cosieves (collections of morphisms out of a fixed object, closed under
postcomposition) in place of sieves.

Rather than developing this dual theory from scratch (repeating every definition,
lemma, and proof of \S\ref{sec:centers}--\S\ref{sec:universal-property} with arrows reversed),
the formalization derives it for free, by definitional transport across the
opposite-category duality $\calC \leftrightsquigarrow \calC^{\op}$.

\begin{fact}
A cosieve from an object $X$ of $\calC$ is the same data as a sieve over $X$, regarded as an
object of $\calC^{\op}$.
\end{fact}

In Lean this identification is recorded as an explicit \lc{Equiv} (\lc{Cosieve.equivSieveOp}),
not merely as a pair of maps going back and forth: a \lcode{Cosieve X} is a structure of
identical shape to \lc{Sieve}, and the equivalence simply reindexes the closure condition
along $\op$.

\begin{definition}[{\citep[Definition~4.1]{Mayeux}}]
A cocenter in $\calC$ is a family $\{(d_i,V_i)\}_{i\in I}$ with $d_i$ a morphism and
$V_i$ a cosieve from $\dom(d_i)$.
\end{definition}
Given the identification above, a cocenter in $\calC$ is a center in $\calC^{\op}$
(with $d_i$ replaced by its formal opposite $d_i^{\op} : \cod(d_i) \to \dom(d_i)$, now going
the right way to serve as a center morphism in $\calC^{\op}$); Lean's
\lc{Cocenter.toCenterOp} records exactly this re-casting, and no new \lc{structure} for
``cocenter data attached to $\calC$ itself'' beyond the re-casting is introduced.

\begin{definition}[{\citep[Definition~4.3]{Mayeux}}]
The codilatation of $\calC$ with cocenter $\{(d_i,V_i)\}_{i\in I}$ is
$\calC[\{V_i\circ(d_i)^{-1}\}_{i\in I}] := \big(\calC^{\op}[\{(d_i)^{-1}\circ
V_i\}_{i\in I}]\big)^{\op}$.
\end{definition}

\begin{leancode}
def Codila (co : Cocenter C) : Type u := (Dila (co.toCenterOp))ᵒᵖ \\
\end{leancode}
This one line is the whole construction: dilate $\calC^{\op}$ by the corresponding
center, then take the opposite category again to land back in (a category built from)
$\calC$. Every structural fact then transports along the same route, with no new proof
content:

\begin{proposition}[{\citep[Proposition~4.5]{Mayeux}}]
There is a canonical functor $\Upsilon : \calC \to \calC[\{V_i\circ(d_i)^{-1}\}_{i\in I}]$,
a canonical faithful functor from the codilatation to $\calC[\{d_i\}_{i\in I}^{-1}]$,
$\Upsilon$ is $\{d_i\}_{i\in I}$-regular, and $\Upsilon$ represents the appropriately dualized
functor $Cat_\calC^{\{d_i\}_{i\in I}\text{-reg}} \to \mathrm{Set}$.
\end{proposition}

In Lean, $\Upsilon$ is \lcode{Cocenter.Upsilon co}, built as \lcode{(CatToDila
co.toCenterOp).rightOp} (Mathlib's operation turning a functor $\calC^{\op} \to \calD$ into a
functor $\calC \to \calD^{\op}$); the faithful comparison functor is
\lc{Codila.faithful_to_loc_op}, an instance derived directly from
Fact~\ref{fact:faithful-into-localization} applied in $\calC^{\op}$; $\Sigma$-regularity of
$\Upsilon$ is \lc{Cocenter.Upsilon_isSigmaRegular}, from Fact~\ref{fact:theta-regular}
transported across $\op$ (using, once more,
\lc{faithful_of_comp_faithful} to relate faithfulness of a functor and of its
\lc{.op}); and representability is \lc{Cocenter.represents}, from
\citep[Proposition~3.12]{Mayeux} transported across $\op$, combined with the sieve/cosieve identification
above. 
\section{Comparison with dilatations of rings}
\label{sec:rings}

The construction that motivated this whole theory, dilatations of a commutative ring $A$
at a multi-center $\{(M_i,a_i)\}_{i\in I}$ of ideals and elements, should be a special case
of dilatations of categories. We present the formalization's own construction first, in the
order the source builds it, and only afterward state the printed paper's Proposition~5.1 and
compare the two: as with \citep[Proposition~3.15]{Mayeux} above, the comparison is easiest to
see
once the construction is already on the table.

\subsection{The construction, following the formalization}
\label{sec:rings-construction}

\subsubsection*{The vendored ring dilatation}
The ring $A[M]$ itself is not developed inside the file discussed in this paper: it is
vendored, essentially verbatim, from the Lean formalization of multi-centered dilatations of
rings~\citep{MZ}, accompanying the separate paper introducing the construction
mathematically~\citep{M}. The construction is recalled, with its Lean rendering, in
\S\ref{sec:appC-construction}; we use it here as a black box. The new content on the ring side is the comparison
with dilatations of categories, described next, together with the elementary properties
collected in Appendix~\ref{app:ring-elementary}.

\subsubsection*{The index type $M^{\mathbb N}$}
\begin{leancode}
def centerOfMulticenter (M : Multicenter A') : Center (CategoryTheory.SingleObj A') where \\
  I := M^ℕ \\
  dom _ := CategoryTheory.SingleObj.star A' \\
  cod _ := CategoryTheory.SingleObj.star A' \\
  mor ν := M.elem^ν \\
  N ν := Sieve.ofIdeal (M.LargeIdeal^ν) \\
\end{leancode}
The index type of this center is $M^{\mathbb N} = \mathbb N^{(I)}$, the same exponent
profiles indexing $A[M]$'s elements above --- \emph{not} $M.\mathrm{index} = I$: the generator at
$\nu$ divides by $a^\nu$, with numerator ranging over all of $L^\nu$.

\subsubsection*{The universal-property half}
Feeding the canonical ring map $A \to A[M]$ through \lc{SingleObj} gives
\lc{toDilatationFunctor}$: $\lc{SingleObj}$\,A \to $\lc{SingleObj}$\,A[M]$, the functor
playing the role of $F$ in Theorem~\ref{thm:universal-property}. \lc{prop_5_1} applies that
theorem to it directly: the sieve inequality is a direct computation (an element of the image
of $L^\nu$ already factors through $a^\nu$, by definition of $L^\nu$), and $\Sigma$-regularity
is built by composing with the further localization of $A[M]$ at the submonoid generated by
the images of the $a_i$'s --- faithful because those images are non-zero-divisors in $A[M]$
(\lc{nonzerodiv_image_single}, an unconditional structural fact about the vendored
construction), so this further localization is injective. This gives
\begin{leancode}
noncomputable def Phi51 : Dila (centerOfMulticenter M) ⥤ CategoryTheory.SingleObj A'[M] := \\
  (prop_5_1 M).choose \\
\end{leancode}
together with \lc{Phi51_spec} and \lc{Phi51_unique}, exactly as \lc{Alpha315} was extracted
from \lc{Dila_universal_property} in \S\ref{sec:combining-construction}.

\subsubsection*{Faithfulness, via a commutativity argument}
Theorem~\ref{thm:universal-property} only guarantees $\Phi=$\lc{Phi51} exists; showing it is
an isomorphism is separate work. Faithfulness goes through a second comparison functor,
\lc{kappaFunctor}$: $\lc{SingleObj}$\,A[M] \to $ the raw localization of
\lc{centerOfMulticenter}$\,M$. Write $S \subset A$ for the multiplicative submonoid generated
by the $a_i$. The monoid map \lc{LocEndLift} extends \lc{toLocEnd}$: A \to $\lc{End}$(\bullet)$
along the localization $A \to S^{-1}A$, and \lc{kappaFunctor} is induced by its composite with
the canonical ring map $A[M] \to S^{-1}A$ (\lc{descToLoc}); the dilatation $A[M]$ itself is not
a localization of $A$ in general. Building
\lc{LocEndLift} needs the target monoid to be commutative, which is not automatic --- it is
established by \lc{allCentral}: every endomorphism of the raw localization's single object
commutes with every other morphism, proved by showing the relevant \lc{MorphismProperty} is
stable under composition, contains every generator-image (central because $A$ is
commutative, \lc{genImage\_central}), and contains every formal inverse (a central element's
inverse is central), hence is everything
(\lc{Localization.Construction.morphismProperty\_is\_top}). With commutativity in hand,
\lc{Phi51}$\ggg$\lc{kappaFunctor} is identified with the (unconditionally faithful, Fact~\ref{fact:faithful-into-localization})
\lc{DilaToLoc} of \lc{centerOfMulticenter}$\,M$ by \lc{Dila\_factor\_unique}, and
\lc{faithful\_of\_comp\_faithful} gives \lc{Phi51\_faithful}.

\subsubsection*{Fullness, for free from the indexing}
Fullness needs none of this machinery. Since the center is indexed by exponent profiles,
every element of $A[M]$ is \emph{directly} a pair $(\nu,m)$ by
\lc{Multicenter.Dilatation.induction\_on}, hence directly the image of the single fraction
generator at $(\nu,m)$, using the defining identity $a^\nu\cdot(m/a^\nu)=m$ inside $A[M]$
itself (\lc{algebraMap\_elem\_pow\_mul\_frac}). No induction on composites is needed, unlike
the fullness arguments of \S\ref{sec:restriction} and \S\ref{sec:counterexample}.

\subsubsection*{Assembling the isomorphism}
With \lc{Phi51} full and faithful, an explicit inverse \lc{Psi51} is built directly (rather
than invoked from a general full-faithful-essentially-surjective equivalence principle, since
both categories here have a single object, so $\Phi$ is precisely a bijective monoid
homomorphism and its set-theoretic inverse is again one), and
\begin{leancode}
noncomputable def Iso51 : Cat.of (Dila (centerOfMulticenter M)) ≅ \\
    Cat.of (CategoryTheory.SingleObj A'[M]) where \\
  hom := (Phi51 M).toCatHom \\
  inv := (Psi51 M).toCatHom \\
  hom_inv_id := Cat.ext <| Phi51_comp_Psi51 M \\
  inv_hom_id := Cat.ext <| Psi51_comp_Phi51 M \\
\end{leancode}
assembles the two composite-identity theorems \lc{Phi51\_comp\_Psi51}/\lc{Psi51\_comp\_Phi51}
into an isomorphism of categories, matching the pattern of \lc{Iso315} in
\S\ref{sec:combining-construction} exactly.

The identification above assembles into a single chain together with one further fact, purely
ring-theoretic and proved independently of categories: reindexing the multi-center by exponent
profiles and dilating again recovers the same ring.

\begin{theorem}
\label{thm:rings-reindex}
Let $\{[M_i,a_i]\}_{i\in I}$ be a multi-center on $A$, with large ideals $L_i = M_i+(a_i)$, and
for $\nu\in\mathbb N^{(I)}$ write $L^\nu := \prod_i L_i^{\nu_i}$, $a^\nu := \prod_i a_i^{\nu_i}$.
For a ring $R$, write $\mathbf B R$ for the one-object category attached to $R$ (its
endomorphism monoid is the multiplicative monoid of $R$), and put $\calC = \mathbf B A$. Then:
\begin{enumerate}
\item[(a)] there is an isomorphism of categories
  $\calC\big[\{(a^\nu)^{-1}\circ L^\nu\}_{\nu\in\mathbb N^{(I)}}\big] \cong \mathbf B\big(A[M]\big)$,
  compatible with the functors from $\calC$;
\item[(b)] there is a ring isomorphism
  $A\big[\{L^\nu/a^\nu\}_{\nu\in\mathbb N^{(I)}}\big] \cong A\big[\{M_i/a_i\}_{i\in I}\big] = A[M]$.
\end{enumerate}
\end{theorem}

Statement (a) is \lc{Iso51} above, with compatibility \lc{Phi51_spec}: the categorical
dilatation by the $\nu$-indexed center \lc{centerOfMulticenter}$\,M$ is isomorphic to
$\mathbf B(A[M])$. Statement (b) is
\lc{Multicenter.Dilatation.reindexRingEquiv}: dilating $A$ by the $\nu$-indexed multi-center
$\{[L^\nu,a^\nu]\}_{\nu\in\mathbb N^{(I)}}$ (\lc{Multicenter.reindex}$\,M$) recovers $A[M]$ up to
isomorphism, via the flattening map $\mu\mapsto\sum_\nu \mu(\nu)\cdot\nu$ sending the reindexed
ring's own exponent profiles back down to $M$'s. 

\subsection{Comparison with the printed paper}
\label{sec:rings-gap}

With the multi-center $\{[M_i,a_i]\}_{i\in I}$ as above, \citep[Proposition~5.1]{Mayeux}
asserts an identification, as $\calC$-categories, of $\calC[\{(a_i)^{-1}\circ M_i\}_{i\in I}]$
(dilating by the center indexed directly by $i\in I$) with the category attached to $A[M]$.
This identification does not hold.

The printed proof's comparison functor $\Phi$ sends a fraction generator to $m/a_i$ and is
claimed surjective. It is not: composition in \lc{SingleObj}$\,A$ is multiplication, so the
image of every morphism is a product $c\cdot\prod_{j=1}^k(m_j/a_{i_j})$ with $c\in A$, $k\ge 0$
and $m_j \in M_{i_j}$, and not every element of $A[M]$ has this multiplicative form. For
instance, take $A=\mathbb Z[X]$, a single index with $a=2$ and $M=(X)$, so that
$A[M] = \mathbb Z[X/2] \subset \mathbb Q[X]$, and consider $(X+2)/2 = 1 + X/2$. A product with
$k = 0$ is an element of $\mathbb Z[X]$, and $1 + X/2 \notin \mathbb Z[X]$. A product with
$k \geq 1$ lies in the ideal $(X/2)\,A[M]$, since $m_j/2 = (X/2)\,p_j$ for $m_j = X p_j$; every
element of this ideal vanishes at $X = 0$, while $1 + X/2$ takes the value $1$.
Moreover no $\calC$-functor can fix this: \lc{toDilatationFunctor} satisfies both hypotheses
of Theorem~\ref{thm:universal-property} for the $I$-indexed center too, so by uniqueness
$\Phi$ is the \emph{only} candidate, and it is not an isomorphism.

This argument is formalized in full: \lc{centerNaive} is the naive $I$-indexed center,
\lc{PhiNaive} the comparison functor built from Theorem~\ref{thm:universal-property}'s
existence clause, \lc{target_elt_not_in_range} proves $(X+2)/2$ is unreachable, and
\lc{no_C_compatible_equiv} concludes that no functor compatible with the two canonical
inclusion functors is an equivalence.

The structural reason is additive closure: $A[M]$'s numerators range over $L^\nu$, a sum of
products of single-ideal elements, but composition alone only ever produces products. The
formalization's \lc{centerOfMulticenter} (\S\ref{sec:rings-construction}) repairs this by
moving the needed sums into the sieves themselves ($L^\nu$ as a single sieve, rather than
$M_i$ generator by generator); with that center, the identification holds
(Theorem~\ref{thm:rings-reindex} above).

\section{A characterization that fails for categories}
\label{sec:counterexample}

This section formalizes \citep[Fact~5.2]{Mayeux}, as Theorem~\ref{thm:no-realizing-center}.

\subsection{The ring-theoretic fact being tested}

For dilatations of commutative rings, every sub-$A$-algebra of a localization
$A[(\{a_i\}_{i\in I})^{-1}]$ arises as the dilatation of $A$ at a suitable multi-center
$\{[M_j,b_j]\}_{j\in J}$ (\citep[\S 2]{M}, recalled in \citep[\S 5.5]{Mayeux}); the index set $J$
and the elements $b_j$ need not be the family $\{a_i\}_{i\in I}$ itself, that is, the family of
denominators may be enlarged.
Theorem~\ref{thm:no-realizing-center} shows, by an explicit finite counterexample, that the
analogous statement fails for dilatations of categories.

\subsection{The counterexample category}

Let $\calC$ have two objects $X,Y$, with $\Hom(X,X) = \{\id_X\}$, $\Hom(Y,Y)=\{\id_Y\}$,
$\Hom(Y,X) = \varnothing$, and $\Hom(X,Y) = \{a,b\}$ two distinct, non-identity, parallel
morphisms; let $\Gamma = \{b\}$.

\begin{leancode}
inductive Obj : Type \\
  | X | Y \\[0.5\baselineskip]
inductive CHom : Obj → Obj → Type \\
  | idX : CHom .X .X \\
  | idY : CHom .Y .Y \\
  | a : CHom .X .Y \\
  | b : CHom .X .Y \\
\end{leancode}
Composition is a four-line pattern match, and associativity holds by \lcode{cases ...
<;> rfl} over the finitely many cases.

Localizing at $\Gamma=\{b\}$ produces infinitely many morphisms $X \to Y$: $b, a,
ab^{-1}a, \dots$ In Lean, the pairwise distinctness of the first three is proved with a
separating invariant: a functor \lc{FSep} to the one-object groupoid on $(\mathbb Z,+)$
(Mathlib's \lcode{SingleObj (Multiplicative ℤ)}), sending $a \mapsto 1$ and $b\mapsto 0$.
The target being a groupoid, \lc{FSep} inverts $\Gamma$ trivially, hence extends to
$\calC[\Gamma^{-1}]$ by the universal property of localization (\S\ref{sec:localization});
the resulting integer is additive along composition and takes the values $0,1,2$ on
$b, a, ab^{-1}a$ (\lc{pairwise_distinct}).

Now let $\calD$ be the subcategory of $\calC[\Gamma^{-1}]$ with the same objects, with
$\Hom_\calD(Y,X)=\varnothing$, $\Hom_\calD(X,X)=\{\id_X\}$, $\Hom_\calD(Y,Y)=\{\id_Y\}$, and
$\Hom_\calD(X,Y) = \{b,a,ab^{-1}a\}$. The evident functors $\calC \to \calD$ and
$\calD \to \calC[\Gamma^{-1}]$ (the inclusion) compose to the localization functor.

\begin{theorem}[{\citep[Fact~5.2]{Mayeux}}]
\label{thm:no-realizing-center}
$\calD \to \calC[\Gamma^{-1}]$ is faithful, but $\calD$ is not isomorphic, as a
$\calC$-category, to $\calC[\{(d_i)^{-1}\circ N_i\}_{i\in I}]$ for any center
$\{(d_i,N_i)\}_{i\in I}$ on $\calC$.
\end{theorem}

Faithfulness of $\calD \to \calC[\Gamma^{-1}]$ is immediate, $\calD$ being literally a
subcategory. 

\subsection{The case analysis}

The proof considers an arbitrary center $\{(d_i,N_i)\}_{i\in I}$ on $\calC$, writes
$\Sigma=\{d_i\}_{i\in I}$, and distinguishes two cases: either (i) $a \in \Sigma$ with
$\id_Y \in N_a$ or $b \in N_a$ (or symmetrically for $b$), in which case
$\#\Hom_{\calC'}(X,Y)$ is infinite; or (ii) whenever $a \in \Sigma$ then $N_a \subset \{a\}$,
and whenever $b\in\Sigma$ then $N_b \subset\{b\}$, in which case $\calC' = \calC$. In both
cases $\calC' \neq \calD$. In Lean, the dichotomy is expressed by a single predicate, with
case~(i) as its negation.

\begin{leancode}
def GoodCenter (Z : Center Obj) : Prop := \\
  ∀ (i : Z.I) (X' : Obj) (m : X' ⟶ Z.cod i), Z.N i m → ∃ q : X' ⟶ Z.dom i, m = q ≫ Z.mor i \\
\end{leancode}
That is: $Z$ is good if every sieve witness factors through its own generator; a case
analysis on the finitely many shapes of generators (\lc{Center.dom_cod_cases}) shows this
is equivalent, for a center on $\calC$, to case~(ii).

\subsubsection{If $Z$ is not good}

Suppose some witness $m \in N_i$ does not factor through $d_i$. Case-splitting on the shape
of $d_i$ (\lc{Center.dom_cod_cases}):
\begin{itemize}
\item if $d_i : X\to X$ or $d_i:Y\to Y$, then $d_i = \id$, through which everything factors,
  contradicting the choice of $m$ (\lc{false_of_hnq_selfmor});
\item if $d_i : X \to Y$, a further case split on $m$ shows that the fraction morphism of
  Proposition~\ref{prop:fraction-exists} for $(i,m)$ is a non-identity endomorphism of $X$
  or a morphism $Y \to X$ in $\calC'$, and $\calD$ has neither (\lc{false_of_hnq_case3}).
\end{itemize}
In Lean, case~(i)'s obstruction is thus exhibited as a single explicit morphism that cannot
exist in $\calD$ under any isomorphism $\calC' \cong \calD$ compatible with the maps from
$\calC$.

\subsubsection{If $Z$ is good}

Suppose $Z$ is good. Lemma~\ref{lem:induction} then shows every morphism of $\calC'$ is
$\Theta$ of a morphism of $\calC$: on a fraction generator $(i,n)$, goodness gives
$n = d_i \circ q$, and cancelling the monomorphism $\Theta(d_i)$
(\citep[Proposition~3.3]{Mayeux}) in the defining triangle forces $b = \Theta(q)$. This
gives \lc{CatToDila_full_of_good}: $\Theta$ is full, and faithful here, so
$\calC' \cong \calC$ as $\calC$-categories; since $\calC$ has two morphisms $X \to Y$ and
$\calD$ has three, $\calC' \not\cong \calD$.

\subsection{Assembling the theorem}

\begin{leancode}
theorem no_realizing_center : \\
    ¬ ∃ (Z : Center Obj) (e : DObj ≌ Dila Z), CtoD ⋙ e.functor = CatToDila Z \\
\end{leancode}
Given a hypothetical center $Z$ and equivalence $e$ realizing $\calD$ as \lc{Dila}$\,Z$
compatibly with the maps from $\calC$, \lc{no_realizing_center} case-splits on
\lc{GoodCenter}: if $Z$ is not good, the impossible morphism of the first case transports
across $e$ to a contradiction; if $Z$ is good, the second case produces an injective
comparison $\psi : \Hom_\calD(X,Y) \to \Hom_{\calC'}(X,Y)$ under which the three generators
$b,a,c$ of $\calD$ (with $c$ standing for $ab^{-1}a$) all land in the two-element set
$\{\Theta(a),\Theta(b)\}$, forcing a collision that contradicts \lc{pairwise_distinct}.

Theorem~\ref{thm:no-realizing-center} is stated for isomorphisms of $\calC$-categories, while
\lc{no_realizing_center} excludes equivalences $e$ with \lcode{CtoD ⋙ e.functor = CatToDila Z}.
The two formulations are equivalent here. An isomorphism is an equivalence. Conversely, both
\lc{CtoD} and \lc{CatToDila Z} are bijective on objects, so an equivalence compatible with them
is bijective on objects, and a fully faithful functor bijective on objects is an isomorphism of
categories.

\appendix
\section{Dictionary between the mathematics and the Lean formalization}
\label{app:dictionary}

This appendix tabulates, for each numbered statement of the original exposition of
dilatations of categories, the Lean declaration(s) that play its role
(\S\ref{sec:dict-notation}--\S\ref{sec:dict-5}); the few statements that are not formalized
are marked ``---'', with the reason in the Comments column.

\subsection{Notation dictionary}
\label{sec:dict-notation}

\begingroup
\renewcommand{\arraystretch}{1.15}
\begin{longtable}{@{}p{2.3cm}p{3.9cm}p{5.1cm}@{}}
\toprule
\textbf{Mathematics} & \textbf{Lean} & \textbf{Comments} \\
\midrule
\endfirsthead
\toprule
\textbf{Mathematics} & \textbf{Lean} & \textbf{Comments} \\
\midrule
\endhead
\bottomrule
\endfoot
$\calC$, $\calD$ & \lc{C}, \lc{D} & \lcode{\{C : Type u\} [Category.\{v\} C]}. \\
$X \xrightarrow{a} Y$ & \lcode{a : X }\ensuremath{\longrightarrow}\lcode{ Y} & \ensuremath{\longrightarrow} is Mathlib's hom-notation. \\
$g \circ f$ & \lcode{f }\ensuremath{\gg}\lcode{ g} & diagrammatic order, see \S\ref{sec:conventions}. \\
$\id_X$ & \lcode{𝟙 X} & \\
sieve over $X$ & \lcode{Sieve X} & Mathlib type, reused as is. \\
cosieve from $X$ & \lcode{Cosieve X} & introduced for this project, \S\ref{sec:dict-4}. \\
$S^{\calC}_E$ & \lc{Sieve.generate} (applied to a \lc{Presieve}) & e.g.\ \lcode{Sieve.generate (Presieve.singleton g)} for $E=\{g\}$. \\
$CoS^{\calC}_E$ & \lc{Cosieve.generate} & \\
$\{[N_i,d_i]\}_{i\in I}$ (\citep[Definition~2.9]{Mayeux}) & \lcode{Center C} & structure, \S\ref{sec:dict-2}. \\
$\{[V_i,d_i]\}_{i\in I}$ & \lcode{Cocenter C} & structure, \S\ref{sec:dict-4}. \\
$\calC[\Sigma^{-1}]$ & \lcode{(Center\allowbreak MorphismProperty Z).\allowbreak Localization} & Mathlib's \lc{MorphismProperty.Localization}. \\
$L : \calC \to \calC[\Sigma^{-1}]$ & \lcode{LocalizationFunctor Z} & \lcode{= (CenterMorphismProperty Z).Q}. \\
$\calC[\{(d_i)^{-1}\circ N_i\}_{i\in I}]$ & \lcode{Dila Z} & \S\ref{sec:dict-2}. \\
$\calC[\{V_i \circ (d_i)^{-1}\}_{i\in I}]$ & \lcode{Codila co} & \S\ref{sec:dict-4}. \\
$\Theta : \calC \to \calC'$ & \lcode{CatToDila Z} & \\
$\Upsilon : \calC \to \calC[\{V_i\circ(d_i)^{-1}\}]$ & \lcode{Cocenter.Upsilon co} & \\
$F'$ (\citep[Theorem~3.10]{Mayeux}) & \lcode{DilaLift Z F hfaith hsieve} & the unique functor with $F' \circ \Theta = F$; defining equation \lc{DilaLift_fac}, uniqueness \lc{DilaLift_unique} (\S\ref{sec:universal-property}). \\
$b = d_i \backslash n = (d_i)^{-1}\circ n$ & \lcode{fraction\allowbreak_in_dila\allowbreak_single Z }\ensuremath{\langle}\lcode{i, }\ensuremath{\langle}\lcode{_, }\ensuremath{\langle}\lcode{n, hn}\ensuremath{\rangle}\ensuremath{\rangle}\ensuremath{\rangle} & the witness of Proposition~\ref{prop:fraction-exists}. \\
$\ell_{d_i} \circ n$ (representative of $b$) & \lcode{fraction\allowbreak_in_path\allowbreak_single Z p}, \lcode{fraction\allowbreak_in_loc\allowbreak_single Z p} & the path in \lcode{Paths (LocQuiver }\ensuremath{\Sigma}\lcode{)}, resp.\ its class in $\calC[\Sigma^{-1}]$; \lcode{inv\allowbreak_in\allowbreak_path Z p} is the $\ell_{d_i}$ factor (Mathlib's \ensuremath{\psi_2}), see \S\ref{def:pair-fraction}. \\
$Cat_\calC^{\Sigma\text{-reg}}$, membership & \lcode{IsSigmaRegular Z F} & a \lc{Prop}, not a category; see \S\ref{sec:sigma-regular}. \\
$S^{\calC'}_{\Theta(N_i)}$, $S^{\calC'}_{\Theta(d_i)}$ & \lcode{CatToDilaSieve Z (Z.N i)}, \lcode{Sieve.generate (Presieve.singleton ((CatToDila Z).map (Z.mor i)))} & \citep[Definition~3.4]{Mayeux} is not given its own name; the two sieves are written out at each use. \\
$A[\{M_i/a_i\}_{i\in I}]$ (ring dilatation) & \lc{A'[M]} (notation for \lc{Multicenter.Dilatation}) & \S\ref{sec:dict-5}. \\
bimorphism & \lcode{Mono f }\ensuremath{\wedge}\lcode{ Epi f} & no dedicated ``Bimorphism'' class; spelled out. \\
\end{longtable}
\endgroup

\subsection{Section 2: sieves, localizations, centers, and dilatations}
\label{sec:dict-2}

As explained in \S\ref{sec:localization}, Definitions~2.1--2.8 (sieves, the graph, $\Sigma$-sequences
and their equivalence, $\Sigma$-fractions, the localization, and its universal property) are
formalized by reusing Mathlib's calculus-of-fractions localization rather than by
reformalizing the graph-theoretic argument of \S\ref{sec:localization} verbatim; the table below
records, for each paper statement, which Mathlib declaration plays its role. From
\citep[Definition~2.9]{Mayeux} on, every declaration listed is specific to this project.

\begingroup
\renewcommand{\arraystretch}{1.15}
\begin{longtable}{@{}p{2.55cm}p{4.1cm}p{4.75cm}@{}}
\toprule
\textbf{Mathematics} & \textbf{Lean declaration} & \textbf{Comments} \\
\midrule
\endfirsthead
\toprule
\textbf{Mathematics} & \textbf{Lean declaration} & \textbf{Comments} \\
\midrule
\endhead
\bottomrule
\endfoot
\citep[Definition~2.1]{Mayeux} (sieve, $S^\calC_E$) & \lc{Sieve}, \lc{Sieve.generate}, \lc{Cosieve}, \lc{Cosieve.generate} & Mathlib's \lc{Sieve}; \lc{Cosieve} is introduced for this project (\S\ref{sec:dict-4}) since Mathlib has no cosieve type. \\
\citep[Definition~2.2]{Mayeux} (graph $\calG$) & \lc{Localization.Construction.LocQuiver} & Mathlib; one edge $a$ per morphism of $\calC$, one edge $l_d$ per $d\in\Sigma$. \\
\citep[Definition~2.3]{Mayeux} ($\Sigma$-sequence) & \lc{Quiver.Path}, \lc{Paths} & Mathlib; paths in the \lc{LocQuiver}. \\
\citep[Definition~2.4]{Mayeux} (equivalence of sequences) & \lc{Localization.Construction.relations} & Mathlib; a congruence on \lcode{Paths (LocQuiver }\ensuremath{\Sigma}\lcode{)}. \\
\citep[Definition~2.5]{Mayeux} ($\Sigma$-fraction) & \lc{MorphismProperty.Localization} & Mathlib; underlying type of \lcode{(CenterMorphismProperty Z).Localization}. \\
\citep[Fact~2.6]{Mayeux} (composite of $l_{d'}, l_d$) & --- & no separate Lean counterpart; a special case of the relations of \citep[Definition~2.4]{Mayeux}/\lc{Localization.Construction.relations}, not restated on its own. \\
\citep[Definition~2.7]{Mayeux} ($\calC[\Sigma^{-1}]$, $L$) & \lc{CenterLocalization}, \lc{LocalizationFunctor} & thin wrappers around \lcode{(CenterMorphismProperty Z).Localization} / \lc{.Q}. \\
\citep[Proposition~2.8]{Mayeux} (universal property of $\calC[\Sigma^{-1}]$) & \lc{Localization.Construction.lift}, \lc{Localization.Construction.fac}, \lc{Localization.Construction.uniq} & Mathlib; recovered as an instance of \citep[Theorem~3.10]{Mayeux} in \lc{Center.ofMorphisms_universal_property}, see \S\ref{sec:dict-5}. \\
\citep[Definition~2.9]{Mayeux} (center) & \lc{Center} & structure with fields \lc{I}, \lc{dom}, \lc{cod}, \lc{mor}, \lc{N}; see \S\ref{sec:centers}. \\
\citep[Definition~2.10]{Mayeux} ($\{[N_i,d_i]\}$-fraction) & \lc{CenterSievePair}, \lc{inv_in_path}, \lc{fraction_in_path_single}, \lc{fraction_in_loc_single}, \lc{IsPairMor}, \lc{PairMorWitness}, \lc{GeneratorMorphismData.fraction} & \lcode{CenterSievePair Z := }\ensuremath{\Sigma}\lcode{ i : Z.I, }\ensuremath{\Sigma}\lcode{ X : C, {f : X }\ensuremath{\longrightarrow}\lcode{ Z.cod i // Z.N i f}} records one $(i,n)$-generator; its underlying morphism is built in three steps --- \lc{inv_in_path} (the edge $\ell_{d_i}$, Mathlib's \ensuremath{\psi_2}), \lc{fraction_in_path_single} (the path-level composite $\ell_{d_i} \circ n$), \lc{fraction_in_loc_single} (its class in $\calC[\Sigma^{-1}]$) --- and \lc{GeneratorMorphismData} then builds the quiver of \S\ref{sec:dilatation-construction}. \\
\citep[Fact~2.11]{Mayeux} (associativity of fraction composition) & \lc{Paths.categoryPaths}, \lc{CategoryTheory.Quotient.category} & the category structure, associativity included, comes from Mathlib's path-category and quotient-category instances; mathematically, associativity is inherited from $\calC[\Sigma^{-1}]$ (\S\ref{sec:fraction-sequences}). \\
\citep[Remark~2.12]{Mayeux} & \lc{GeneratorMorphismData.original} & the two clauses of the remark are definitional consequences of how \lc{GeneratorQuiver} is built, not separate lemmas. \\
\citep[Definition~2.13]{Mayeux} ($\calC'=\calC[\{(d_i)^{-1}\circ N_i\}]$) & \lc{Dila}, \lc{DilaRel}, \lc{GeneratedCategory}, \lc{GeneratedToDila} & $\calC'$ \lcode{:= Quotient (DilaRel Z)}; see \S\ref{sec:dilatation-construction}. \\
\citep[Fact~2.14]{Mayeux} (faithful functor $\calC' \to \calC[\Sigma^{-1}]$) & \lc{DilaToLoc}, \lc{Fact_2_14}, \lc{DilaGenerators}, \lc{DilaToLoc_range} & \lcode{Fact_2_14 : (DilaToLoc Z).Faithful}; \lc{DilaToLoc_range}: its image is the subcategory generated by \lc{DilaGenerators} (\S\ref{sec:dilatation-construction}). \\
\citep[Fact~2.15]{Mayeux} ($\calC' \cong \calC[\Sigma^{-1}]$ when all $N_i = S^\calC_{Id_{cod(d_i)}}$) & \lc{Center.ofMorphisms}, \lc{CatToDila_ofMorphisms_isIso}, \lc{Fact_2_15} & \lc{Center.ofMorphisms} builds the center with \lcode{N := fun _ => }\ensuremath{\top}; \lc{Fact_2_15} is the isomorphism \lcode{Cat.of (Dila _) }\ensuremath{\cong}\lcode{ Cat.of (... .Localization)}. \\
\citep[Corollary~2.16]{Mayeux} (smallness) & --- & not formalized; size discipline is handled by \lcode{Center C}'s universe annotations rather than an explicit smallness lemma. \\
\end{longtable}
\endgroup

\subsection{Section 3: the universal property and related results}
\label{sec:dict-3}

\begingroup
\renewcommand{\arraystretch}{1.15}
\begin{longtable}{@{}p{2.55cm}p{4.1cm}p{4.75cm}@{}}
\toprule
\textbf{Mathematics} & \textbf{Lean declaration} & \textbf{Comments} \\
\midrule
\endfirsthead
\toprule
\textbf{Mathematics} & \textbf{Lean declaration} & \textbf{Comments} \\
\midrule
\endhead
\bottomrule
\endfoot
\citep[Proposition~3.1(i)]{Mayeux} ($\Theta : \calC \to \calC'$) & \lc{CatToDila} & \\
\citep[Proposition~3.1(ii)]{Mayeux} (fraction $b=d_i\backslash n$) & \lc{fraction_in_dila_single}, \lc{fraction_in_dila_comp_mor}, \lc{fraction_in_dila_unique} & existence and the defining triangle; uniqueness from $\Theta(d_i)$ being a monomorphism (\lc{Prop_3_3}). \\
\citep[Fact~3.2]{Mayeux} (bimorphism under a faithful functor) & \lc{Fact_3_2} & stated for arbitrary \lcode{F : A }\ensuremath{\Rightarrow}\lcode{ B} with \lc{[F.Faithful]}, not specialized to $\Theta$. \\
\citep[Proposition~3.3]{Mayeux} ($\Theta(d_i)$ is a bimorphism) & \lc{Prop_3_3} & \\
\citep[Definition~3.4]{Mayeux} ($S^{\calC'}_{\Theta(N_i)}$, $S^{\calC'}_{\Theta(d_i)}$) & \lc{CatToDilaSieve} & \lcode{CatToDilaSieve Z (Z.N i) := Sieve.functorPushforward (CatToDila Z) (Z.N i)}; $S^{\calC'}_{\Theta(d_i)}$ itself is not named and is written out as \lcode{Sieve.generate (Presieve.singleton ((CatToDila Z).map (Z.mor i)))} at each use. \\
\citep[Proposition~3.5]{Mayeux} ($S^{\calC'}_{\Theta(N_i)} \subset S^{\calC'}_{\Theta(d_i)}$) & \lc{CatToDila_image_sieve_le_singleton} & \\
\citep[Definition~3.6]{Mayeux} ($Cat_\calC^{\Sigma\text{-reg}}$) & \lc{IsSigmaRegular} & a \lc{Prop} (membership only), not a category; see \S\ref{sec:sigma-regular}. \\
\citep[Fact~3.7]{Mayeux} ($\Theta$ is $\Sigma$-regular) & \lc{CatToDila_isSigmaRegular} & \\
\citep[Fact~3.8]{Mayeux} ($\Sigma'\subset\Sigma$, $L$ faithful $\Rightarrow L'$ faithful) & \lc{faithful_of_comp_faithful}, \lc{faithful_of_comp_faithful_gen} & replaces the paper's triangle-of-functors argument with the purely categorical fact ``if $p\ggg e$ is faithful then $p$ is faithful,'' which also drives \citep[Fact~3.7]{Mayeux}; applied instance: \lc{IsSigmaRegular_sum_inl_of} (\citep[Proposition~3.15]{Mayeux}). \\
\citep[Remark~3.9]{Mayeux} (Buan--Marsh) & --- & not formalized, and not needed: the paper itself states it is not used in the rest of the text. \\
\citep[Theorem~3.10]{Mayeux} (universal property) & \lc{DilaLift} ($F'$), \lc{DilaLift_fac}, \lc{DilaLift_unique}, \lc{Dila_universal_property} & \lc{DilaLift} is $F'$ as a named construction (a quotient-lift of the path-lift $H$, \S\ref{sec:universal-property}); \lc{DilaLift_fac} is $F' \circ \Theta = F$, \lc{DilaLift_unique} its uniqueness, and \lc{Dila_universal_property} their packaging as \ensuremath{\exists}\lcode{! (G : Dila Z }\ensuremath{\Rightarrow}\lcode{ D), CatToDila Z }\ensuremath{\ggg}\lcode{ G = F}, under hypotheses \lcode{hfaith : (ImageCenterLocalizationFunctor Z F).Faithful} and \lcode{hsieve : }\ensuremath{\forall}\lcode{ i, Sieve.functorPushforward F (Z.N i) }\ensuremath{\leq}\lcode{ Sieve.generate (Presieve.singleton (F.map (Z.mor i)))}, i.e.\ conditions~(2),(1) kept separate rather than bundled as membership in $Cat_\calC^{\Sigma\text{-reg}}$; see \S\ref{sec:sigma-regular}. \\
\citep[Fact~3.11]{Mayeux} (pushforward of a generated sieve) & \lc{CatToDila_comp_image_sieve_le_singleton} & the composite-functor form used directly in the proof of \citep[Theorem~3.10]{Mayeux} and \citep[Proposition~3.12]{Mayeux}; the bare statement about an arbitrary functor $H$ is Mathlib's \lc{Sieve.functorPushforward_comp}. \\
\citep[Proposition~3.12]{Mayeux} ($\Theta$ represents $Cat_\calC^{\Sigma\text{-reg}}\to Set$) & \lc{CatToDila_represents} & stated as an ``iff'': \lcode{(}\ensuremath{\exists}\lcode{! G, CatToDila Z }\ensuremath{\ggg}\lcode{ G = F) }\ensuremath{\leftrightarrow}\lcode{ }\ensuremath{\forall}\lcode{ i, ...}, rather than as a bijection of \lc{Hom}-sets; equivalent, more convenient to use. \\
\citep[Fact~3.13]{Mayeux} (subsieve $M_i \subset N_i$ gives a faithful $\varphi$) & \lc{Fact313Phi}, \lc{IsSigmaRegular_altSieve} & built from \citep[Theorem~3.10]{Mayeux} applied to $\Theta$, exactly as \lc{restrictPhi}; faithfulness via \lc{Dila_factor_unique} and \lc{faithful_of_comp_faithful} (\S\ref{sec:subsieve}). \\
\citep[Proposition~3.14]{Mayeux} ($K \subset I$, functor $\Phi$; full/faithful) & \lc{Center.restrict}, \lc{restrictPhi}, \lc{restrictPhi_full}, \lc{restrictPhi_faithful} & part (i) is \lc{restrictPhi_full} (hypothesis \lcode{hI : }\ensuremath{\forall}\lcode{ i }\ensuremath{\notin}\lcode{ K, Z.N i = Sieve.generate (Presieve.singleton (Z.mor i))}); part (ii) is \lc{restrictPhi_faithful} (hypothesis \lcode{(baseRestrictFunctor Z K).Faithful}). \\
\citep[Proposition~3.15]{Mayeux} (combining two centers) & \lc{Center.sum}, \lc{CenterZW}, \lc{Phi315} ($\Phi$), \lc{Beta315} ($\beta$), \lc{Alpha315} ($\alpha$), \lc{Alpha'315} ($\alpha'$), \lc{Iso315} & (1) \lc{Phi315_isSigmaRegular}; (2) false in general, taken as the hypothesis \lc{hreg} (\S\ref{sec:gap}); (3), (5), (6) under \lc{hreg}: \lc{Alpha315}/\lc{Alpha315_spec}/\lc{Alpha315_unique}, \lc{Phi315_comp_Alpha315}/\lc{Alpha315_comp_Alpha'315}/\lc{Alpha'315_comp_Alpha315}, \lc{Iso315}; (4) \lc{Alpha'315}/\lc{Alpha'315_spec}; counterexample to (2), (3), (5), (6): \lc{Ex315.not_hreg}, \lc{Ex315.no_alpha}, \lc{Ex315.alpha'_not_faithful}, \lc{Ex315.not_equivalent} (\S\ref{sec:counterexample-315}). \\
\citep[Remark~3.16]{Mayeux} (direct computation for 3.15(6)) & --- & not formalized; an alternative proof strategy the paper explicitly does not use either. \\
\citep[Fact~3.17]{Mayeux} (union of pushforward sieves) & \lc{Sieve.functorPushforward_union} & Mathlib. \\
\citep[Proposition~3.18]{Mayeux} ($N_i''=N_i\cup N_i'$) & \lc{Center.sieveUnion}, \lc{Center.altSieve}, \lc{Alpha318}, \lc{Alpha'318}, \lc{Iso318} & mirrors \lc{Alpha315}/\lc{Alpha'315}/\lc{Iso315}, without the extra \lc{hreg} hypothesis (regularity holds unconditionally here). \\
\end{longtable}
\endgroup

\subsection{Section 4: codilatations}
\label{sec:dict-4}

Codilatations are formalized exactly as the paper constructs them: by passing to $\calC^{\op}$,
forming a dilatation there, and passing back; every declaration below is a thin wrapper around
the corresponding dilatation declaration of \S\ref{sec:dict-2}--\ref{sec:dict-3}, composed with
Mathlib's opposite-category API (\lc{Functor.op}, \lc{Functor.rightOp}, \lc{Opposite.op}).

\begingroup
\renewcommand{\arraystretch}{1.15}
\begin{longtable}{@{}p{2.55cm}p{4.1cm}p{4.75cm}@{}}
\toprule
\textbf{Mathematics} & \textbf{Lean declaration} & \textbf{Comments} \\
\midrule
\endfirsthead
\toprule
\textbf{Mathematics} & \textbf{Lean declaration} & \textbf{Comments} \\
\midrule
\endhead
\bottomrule
\endfoot
\citep[Definition~4.1]{Mayeux} (cocenter $\{[V_i,d_i]\}_{i\in I}$) & \lc{Cosieve}, \lc{Cocenter} & \lcode{Cosieve X} mirrors \lc{Sieve} but is stable under postcomposition; \lc{Cocenter} records \lcode{V i : Sieve (Opposite.op (dom i))} directly, already through the identification of \citep[Fact~4.2]{Mayeux}, avoiding re-deriving it at every use. \\
\citep[Fact~4.2]{Mayeux} (cosieve from $X$ = sieve over $\mathrm{op}\,X$ in $\calC^{\op}$) & \lc{Cosieve.toSieveOp}, \lc{Sieve.toCosieveUnop}, \lc{Cosieve.equivSieveOp}, \lc{Cocenter.toCenterOp} & an explicit \lc{Equiv} \lc{Cosieve.equivSieveOp}, not two one-directional maps; \lcode{Cocenter.toCenterOp co : Center C\textsuperscript{op}} bundles the cocenter under this identification, ready to feed into \lc{Dila}. \\
\citep[Definition~4.3]{Mayeux} (codilatation $\calC[\{V_i\circ(d_i)^{-1}\}]$) & \lc{Codila} & \lcode{Codila co := (Dila (co.toCenterOp))\textsuperscript{op}}, verbatim the paper's formula. \\
\citep[Fact~4.4]{Mayeux} ($CoS^\calC_E = S^{\calC^{op}}_E$) & \lc{Cosieve.generate_toSieveOp} & \\
\citep[Proposition~4.5(i)]{Mayeux} ($\Upsilon : \calC \to \calC[\{V_i\circ(d_i)^{-1}\}]$) & \lc{Cocenter.Upsilon} & \lcode{:= (CatToDila (co.toCenterOp)).rightOp}. \\
\citep[Proposition~4.5(ii)]{Mayeux} (faithful $\calC[\{V_i\circ(d_i)^{-1}\}] \to \calC[\{d_i\}^{-1}]$) & \lc{Codila.faithful_to_loc_op} & an instance, obtained directly from \lc{Fact_2_14} applied in $\calC^{op}$. \\
\citep[Proposition~4.5(iii)]{Mayeux} ($\Upsilon \in Cat_\calC^{\{d_i\}\text{-reg}}$) & \lc{Cocenter.Upsilon_isSigmaRegular} & \\
\citep[Proposition~4.5(iv)]{Mayeux} ($\Upsilon$ represents) & \lc{Cocenter.represents} & obtained from \lc{CatToDila_represents} (\citep[Proposition~3.12]{Mayeux}) applied in $\calC^{op}$, combined with \citep[Fact~4.4]{Mayeux}. \\
\end{longtable}
\endgroup

\subsection{Section 5: examples}
\label{sec:dict-5}

Rows follow the paper's order: \S5.1 (the universal property of localizations recovered
from \citep[Theorem~3.10]{Mayeux}), \S5.2 (dilatations of commutative rings,
\citep[Proposition~5.1]{Mayeux}, with the ring-level dilatation vendored from \citep{M}'s
own Lean development), \S5.3--5.4, and \S5.5 (Theorem~\ref{thm:no-realizing-center}).

\begingroup
\renewcommand{\arraystretch}{1.15}
\begin{longtable}{@{}p{2.55cm}p{4.1cm}p{4.75cm}@{}}
\toprule
\textbf{Mathematics} & \textbf{Lean declaration} & \textbf{Comments} \\
\midrule
\endfirsthead
\toprule
\textbf{Mathematics} & \textbf{Lean declaration} & \textbf{Comments} \\
\midrule
\endhead
\bottomrule
\endfoot
\S5.1 (universal property of localizations recovered) & \lc{Center.ofMorphisms}, \lc{Center.ofMorphisms_universal_property}, \lc{Fact_2_15} & the center with all sieves trivial (\lcode{N := fun _ => }\ensuremath{\top}); any $F$ inverting every $d_i$ factors uniquely, i.e.\ \citep[Proposition~2.8]{Mayeux} as a special case of \citep[Theorem~3.10]{Mayeux}. \\
$\{[M_i,a_i]\}_{i\in I}$ (ring center, \citep{M}) & \lcode{Multicenter A'} & vendored from \citep{M}'s own formalization. \\
$A[\{M_i/a_i\}_{i\in I}]$ & \lc{A'[M]} (\lc{Multicenter.Dilatation}) & \\
$\calC$ attached to $A$ (single object) & \lcode{SingleObj A'} & Mathlib. \\
center in $\calC$ attached to $\{[M_i,a_i]\}$ & \lc{centerOfMulticenter} & builds a \lcode{Center (SingleObj A')} from a \lcode{Multicenter A'}. \\
functor $\calC \to \calC[\{M_i/a_i\}]$ & \lc{toDilatationFunctor} & \\
\citep[Proposition~5.1]{Mayeux} & \lc{prop_5_1}, \lc{Iso51} & \lc{prop_5_1} is the universal-property half, via \citep[Theorem~3.10]{Mayeux} (faithfulness from the generators of $M$ being non-zero-divisors in $A'[M]$); \lc{Iso51} assembles it with its inverse \lc{Psi51} into \lcode{Dila (centerOfMulticenter M) }\ensuremath{\cong}\lcode{ SingleObj A'[M]}. \\
naive $\calC[\{(a_i)^{-1}\circ M_i\}_{i\in I}] \cong A[M]$ (false, \S\ref{sec:rings-gap}) & \lc{centerNaive}, \lc{PhiNaive}, \lc{target_elt_not_in_range}, \lc{no_C_compatible_equiv} & \lc{centerNaive} is the literal $i$-indexed center; \lc{PhiNaive} the comparison functor from \citep[Theorem~3.10]{Mayeux}; \lc{target_elt_not_in_range} shows $(X{+}2)/2$ is unreachable; \lc{no_C_compatible_equiv} concludes no compatible functor is an equivalence. \\
\S5.3 (monoid dilatations), \S5.4 (pre-additive categories) & --- & no numbered statements; not formalized as standalone constructions. \\
$\calC$ ($X,Y$, $Hom(X,Y)=\{a,b\}$) (\S5.5) & \lc{Obj}, \lc{CHom}, \lc{CHom.comp} & two-constructor \lcode{inductive Obj | X | Y}, and \lcode{CHom : Obj }\ensuremath{\to}\lcode{ Obj }\ensuremath{\to}\lcode{ Type} with constructors \lcode{idX, idY, a, b}; composition by pattern-matching, associativity by \lcode{cases ... <;> rfl}. \\
$\Gamma=\{b\}$ & \lc{Gamma} & \lcode{MorphismProperty Obj}. \\
$\calC[\Gamma^{-1}]$, distinctness of $b,a,ab^{-1}a$ & \lc{FSep}, \lc{FSep'}, \lc{cMor}, \lc{pairwise_distinct} & separated by a functor to \lcode{SingleObj (Multiplicative }\ensuremath{\mathbb{Z}}\lcode{)} sending $a\mapsto1$, $b\mapsto0$: witnesses the infinitely many morphisms $X\to Y$ and gives a separating invariant for the three named ones. \\
$\calD$ ($Hom(X,Y)=\{b,a,ab^{-1}a\}$) & \lc{DObj}, \lc{DHom}, \lc{DHom.comp} & same pattern as \lc{Obj}/\lc{CHom}, with a third generator \lc{c} standing for $ab^{-1}a$. \\
functors $\calC\to\calD$, $\calD\to\calC[\Gamma^{-1}]$ & \lc{CtoD}, \lc{DtoLoc}, \lc{CtoD_comp_DtoLoc} & identity on objects, as in the paper; \lc{CtoD_comp_DtoLoc} is the commuting triangle. \\
faithfulness of $\calD\to\calC[\Gamma^{-1}]$ & \lc{DtoLoc_faithful} & \\
``$\calC' \ne \calD$ for every center'' & \lc{GoodCenter}, \lc{false_of_not_good}, \lc{exists_C_mor_of_good}, \lc{CatToDila_full_of_good}, \lc{no_realizing_center} & see \S\ref{sec:counterexample} for the two-case analysis. \\
\end{longtable}
\endgroup

\section{Erratum to \citep{Mayeux}}
\label{app:erratum}

\begin{itemize}
\item The identification $\calC[\{(a_i)^{-1}\circ M_i\}_{i\in I}] \cong A[M]$ of
\citep[Proposition~5.1]{Mayeux} is false in general (\S\ref{sec:rings-gap}). It should be
replaced by the identification
$\calC[\{(a^\nu)^{-1}\circ L^\nu\}_{\nu\in\mathbb N^{(I)}}] \cong \mathbf B(A[M])$ of
Theorem~\ref{thm:rings-reindex}(a), together with the ring isomorphism
$A[\{L^\nu/a^\nu\}_{\nu\in\mathbb N^{(I)}}] \cong A[\{M_i/a_i\}_{i\in I}] = A[M]$ of
Theorem~\ref{thm:rings-reindex}(b); see \S\ref{sec:rings-construction} above.
\item Part~(2) of \citep[Proposition~3.15]{Mayeux} is false in general
(Proposition~\ref{prop:counterexample-315}), and therefore also parts~(3), (5) and~(6),
which depend on it. Parts~(3), (5) and~(6) hold under the additional hypothesis that
$\Theta_Z\ggg\beta$ is $\{d_k\}_{k\in I'}$-regular, which is part~(2) itself (\lc{hreg},
\S\ref{sec:gap}); parts~(1) and (4) hold unconditionally. See \S\ref{sec:gap} and
\S\ref{sec:counterexample-315} above.
\end{itemize}
The other numbered statements of \citep{Mayeux} listed in Appendix~\ref{app:dictionary} are
formalized as stated, except those marked there as not formalized.

\section{Dilatations of rings}
\label{app:ring-elementary}

The Lean development also contains a section, \lc{AppendixC} in the source, formalizing
material from \citep[\S 2.2]{M} on \emph{dilatations of rings}, independent of the
categorical theory of the body of this paper. The full formalization of \citep{M} will be the topic of a future paper; we record
here only what is relevant to the present one.

\subsection{The construction}
\label{sec:appC-construction}

A \emph{multi-center} in a commutative ring $A$ is a family $\{[M_i,a_i]\}_{i\in I}$ of an
ideal $M_i$ and an element $a_i$ of $A$ for each $i \in I$. In Lean it is a triple of an
index type and two functions:
\begin{leancode}
variable (A' : Type*) [CommSemiring A'] \\[0.5\baselineskip]
structure Multicenter where \\
  (index : Type*) \\
  (ideal : index → Ideal A') \\
  (elem : index → A') \\
\end{leancode}
Write $L_i := M_i+(a_i)$ for the associated \emph{large ideal} (\lc{LargeIdeal}) and, for a
finitely-supported exponent tuple $\nu \in \mathbb N^{(I)}$ (Lean notation $M^{\mathbb N}$
for \lcode{index →₀ ℕ}), $a^\nu := \prod_i a_i^{\nu_i}$ and $L^\nu := \prod_i L_i^{\nu_i}$
(\lc{familyPow}, \lc{prodLargeIdealPower}). A general element of the dilatation
$A[\{M_i/a_i\}_{i\in I}]$ is then a pair $(\nu,m)$ with $m \in L^\nu$ (\lc{PreDil}), modulo
the equivalence $(\nu,m)\sim(\nu',m')$ iff $m\cdot a^{\nu'+\beta} = m'\cdot a^{\nu+\beta}$
for some $\beta\in\mathbb N^{(I)}$ (\lc{Multicenter.r}); the dilatation $A[M]$
(\lc{Multicenter.Dilatation}, notation \lc{A'[M]}) is the resulting quotient. See
\S\ref{sec:rings-construction} for the provenance of this construction and for the use made
of it in the body of the paper.

\subsection{Restricting a multicenter, and the two-stage dilatation}
\label{sec:appC-restrict}

For a subset $K \subset I$ we write $M|_K$ for the multicenter $\{[M_i,a_i]\}_{i\in K}$
obtained by restricting the index set, formalized as \lc{Multicenter.restrict}. Writing
$B = A[\{M_i/a_i\}_{i\in K}]$ for the corresponding first-stage dilatation, the
complementary indices $j \in I \setminus K$ determine a multicenter in $B$, namely
$\{[B\cdot M_j, a_j/1]\}_{j \in I \setminus K}$, formalized as
\lc{Multicenter.complement}; here $B \cdot M_j$ is the ideal of $B$ generated by the image
of $M_j$, rendered in Lean as \lcode{Ideal.map (algebraMap A' B) (M.ideal j)}.

\begin{proposition}[{\citep[Proposition~2.24]{M}}]
\label{prop:app-2-24}
Let $K \subset I$. There is a canonical isomorphism of $A[\{M_i/a_i\}_{i\in K}]$-algebras
\[
  A\Big[\big\{\tfrac{M_i}{a_i}\big\}_{i\in K}\Big]
    \Big[\Big\{\tfrac{A[\{M_i/a_i\}_{i\in K}]\,M_j}{a_j}\Big\}_{j \in I\setminus K}\Big]
  \;\xrightarrow{\ \sim\ }\;
  A\Big[\big\{\tfrac{M_i}{a_i}\big\}_{i\in I}\Big].
\]
In words: dilating first along the indices in $K$ and then along the remaining indices
recovers the dilatation along all of $I$ in one step.
\end{proposition}

The isomorphism of Proposition~\ref{prop:app-2-24} is moreover the \emph{only} one, and this
too is a consequence of the universal property rather than of any computation.

\begin{lemma}[\lc{prop_2_24_unique}, \lc{prop_2_24_existsUnique}]
\label{lem:app-2-24-unique}
With the notation of Proposition~\ref{prop:app-2-24}, there is exactly one isomorphism
\[
  A\Big[\big\{\tfrac{M_i}{a_i}\big\}_{i\in K}\Big]
    \Big[\Big\{\tfrac{A[\{M_i/a_i\}_{i\in K}]\,M_j}{a_j}\Big\}_{j \in I\setminus K}\Big]
  \;\xrightarrow{\ \sim\ }\;
  A\Big[\big\{\tfrac{M_i}{a_i}\big\}_{i\in I}\Big]
\]
compatible with the canonical maps out of $A[\{M_i/a_i\}_{i\in K}]$, namely the one of
Proposition~\ref{prop:app-2-24}.
\end{lemma}

Indeed, the uniqueness half of \citep[Proposition~2.29]{M} already gives more: \emph{every}
$A[\{M_i/a_i\}_{i\in K}]$-algebra map from the two-stage dilatation to
$A[\{M_i/a_i\}_{i\in I}]$ equals the morphism of Proposition~\ref{prop:app-2-24}
(\lc{chi224_unique}).
Uniqueness of the isomorphism follows at once, since an isomorphism of algebras is determined
by its underlying algebra map.

In the Lean source these are \lc{prop_2_24}, \lc{chi224_unique} and
\lc{prop_2_24_existsUnique}; all three are obtained from the universal property of dilatations of
rings (\citep[Proposition~2.29]{M}, in the development \lc{Multicenter.desc} with its
uniqueness statement \lc{Multicenter.lemma_exists_unique_morphism}), applied to the
multicenter \lc{Multicenter.complement}. We do not reproduce the arguments here: they are
those of \citep[\S 2.2]{M}, and the full formalization of \citep{M} is the subject of a
future paper.

\medskip\noindent{\bfseries Statements and declarations.}

\smallskip\noindent{\itshape Competing interests.} The author has no competing interests
to declare.

\smallskip\noindent{\itshape Data availability.} The Lean~4 source code underlying this
work is openly available at \url{https://github.com/rndmx/DilCat}.

\smallskip\noindent This work was carried out during the 2025--2026 academic year.

\medskip\noindent{\bfseries AI use disclosure.}
Claude (Anthropic) assisted with this matter, under the author's direction. In particular, it helped to
detect the error in the printed paper \citep{Mayeux}.


\begin{thebibliography}{99}

\bibitem[Dubouloz, Mayeux and dos Santos(to appear)]{DMS} A.~Dubouloz, A.~Mayeux and
  J.P.~dos Santos, A survey on algebraic dilatations, {\it Fields Institute Monographs},
  Springer, to appear, arXiv:2306.17003.

\bibitem[Gabriel and Zisman(1967)]{GZ} P.~Gabriel and M.~Zisman, {\it Calculus of fractions
  and homotopy theory}, Ergebnisse der Mathematik und ihrer Grenzgebiete 35, Springer-Verlag,
  New York, 1967.

\bibitem[The mathlib Community(2020)]{mathlib} The mathlib Community, The Lean mathematical
  library, in {\it Proceedings of the 9th ACM SIGPLAN International Conference on Certified
  Programs and Proofs (CPP 2020)}, ACM, 2020, 367--381,
  \url{https://doi.org/10.1145/3372885.3373824}.

\bibitem[Mayeux(2026)]{M} A.~Mayeux, Multi-centered dilatations, congruent isomorphisms and
  Rost double deformation space, {\it Transformation Groups}, 31 (2026), 1801--1850,
  \url{https://doi.org/10.1007/s00031-024-09894-9}.

\bibitem[Mayeux(2025)]{Mayeux} A.~Mayeux, Dilatations of categories, {\it Higher Structures},
  9(2), 2025, 62--75, \url{https://doi.org/10.21136/HS.2025.10}.

\bibitem[Mayeux(2026)]{MayDil} A.~Mayeux, Formalizing all indexed mathematics as a
  benchmark for general reasoning: implementing dilatations of categories, in {\it Intelligent
  Systems and Applications --- Proceedings of the 2026 Intelligent Systems Conference
  (IntelliSys)}, Lecture Notes in Networks and Systems 2146, Springer, Cham, 2026, 393--405,
  \url{https://doi.org/10.1007/978-3-032-35078-7_23}.

\bibitem[Mayeux, Richarz and Romagny(2023)]{MRR} A.~Mayeux, T.~Richarz and M.~Romagny,
  N\'eron blowups and low-degree cohomological applications, {\it Algebraic Geometry},
  10(6), 2023, 729--753, \url{https://doi.org/10.14231/AG-2023-026}.

\bibitem[Mayeux and Zhang(2026)]{MZ} A.~Mayeux and J.~Zhang, Formalizing multi-graded
  Brenner--Schr\"oer Proj schemes and dilatations of rings in Lean4, {\it Journal of
  Automated Reasoning}, 70(2), 2026, Article 17,
  \url{https://doi.org/10.1007/s10817-026-09764-x}.

\bibitem[de Moura and Ullrich(2021)]{Lean4} L.~de Moura and S.~Ullrich, The Lean~4 theorem
  prover and programming language, in {\it Automated Deduction --- CADE 28}, Lecture Notes
  in Computer Science 12699, Springer, 2021, 625--635,
  \url{https://doi.org/10.1007/978-3-030-79876-5_37}.

\end{thebibliography}
\end{document}